\documentclass[preprint,12pt]{elsarticle}

\usepackage{amssymb}

\usepackage{lineno}

\usepackage{graphicx}
\usepackage{amsmath}

\journal{}

\begin{document}

\begin{frontmatter}



\title{Cooling rates and intensity limitations for laser-cooled ions at relativistic energies}


\author[TEMF]{Lewin Eidam}
\ead{eidam@temf.tu-darmstadt.de}
\author[TEMF,GSI]{Oliver Boine-Frankenheim}
\author[GSI]{Danyal Winters}

\address[TEMF]{Institut f\"ur Theorie elektromagnetischer Felder, Technical University Darmstadt, Schlossgartenstr. 8 D-64289 Darmstadt, Germany}
\address[GSI]{GSI Helmholtzzentrum f\"ur Schwerionenforschung, Planckstr. 1, D-64291 Darmstadt, Germany}

\begin{abstract}
The ability of laser cooling for relativistic ion beams is investigated. For this purpose, the excitation of relativistic ions with a continuous wave and a pulsed laser is analyzed, utilizing the optical Bloch equations. 
The laser cooling force is derived in detail and its scaling with the relativistic factor $\gamma$ is discussed. The cooling processes with a continuous wave and a pulsed laser system are investigated. 
Optimized cooling scenarios and times are obtained in order to determine the required properties of the laser and the ion beam for the planed experiments. The impact of beam intensity effects, like intrabeam scattering and space charge are analyzed. Predictions from simplified models are compared to particle-in-cell simulations and are found to be in good agreement. Finally two realistic example cases of Carbon ions in the ESR and relativistic Titanium ions in SIS100 are compared in order to discuss prospects for future laser cooling experiments.
\end{abstract}


\end{frontmatter}


\section{Introduction}\label{S:1}

The relative momentum spread of a particle ensemble can be strongly reduced by means of laser cooling \cite{Phillips1998}. This offers great perspectives for future facilities that will provide high-quality stored ion beams for precision experiments. Laser cooling of stored coasting and bunched ion beams has been demonstrated at the TSR in Heidelberg (Germany) \cite{schroeder1990, Lauer1998}, and at ASTRID in Aarhus (Denmark) \cite{Hangst1991}. At the Experimental Storage Ring (ESR) in Darmstadt (Germany), first laser cooling experiments at moderately relativistic energies were conducted \cite{Schramm2005}. Particle intensity limitations due to space charge effects and intrabeam scattering for low energy ion beams were investigated in ASTRID \cite{Hangst1995}.

In the future, laser cooling of intense highly charged ion beams at relativistic energies will be attempted for the first time at the Facility for Antiproton and Ion Research (FAIR), which is presently under construction \cite{Winters2015}. In the heavy ion synchrotron SIS100 only laser cooling will be available, because electron cooling becomes less effective at high relativistic factors ($\gamma \gg 1$) \cite{poth1990}, and stochastic cooling works best at much lower intensities. In this paper the efficiency of laser cooling at high relativistic factors ($1.1\lesssim \gamma \lesssim 12$) will be investigated using analytical as well as numerical models. The results are compared to laser cooling at moderately relativistic factors ($\gamma \lesssim 1.1$) which was experimentally studied in the ESR. This work concentrates on the effects during the cooling process and does not investigate the equilibrium state of ultra cold beams and strongly coupled ensembles (for more information see ref. \cite{schramm2004, noda2014}).

The principle of laser cooling relies on the directional absorption of energy and momentum of resonant laser photons by an ion, and the subsequent random emission of fluorescence photons from the ion and the corresponding randomly distributed recoil momentum. For the laser light to be resonant with a fast atomic transition in the ion, the photon energy needs to be equal to the transition energy. This constraint demands a relationship between the ion type (element, charge state), the wavelength of the laser system and the speed of the ions, which is well described in ref. \cite{Schramm2005}. 
Consequently, a change of the relativistic factor $\gamma$ involves a change of the laser wavelength or the ion type. Due to practical reasons the laser wavelength is restricted to two values. The ion type can be changed over a broad range, of which we show two examples in this investigation. The $2s_{1/2}\rightarrow 2p_{1/2}$ transition in Li-like ions is chosen (values given in ref. \cite{johnson1996}), but the formulas can be applied to any other atomic transition with a lifetime that is short compared to the revolution time of the ion in the accelerator. The particle dynamics are studied for ion energies below the transition energy of the synchrotron ($\gamma < \gamma_t$). The structure of this paper is as follows: Firstly the expected laser forces for continuous wave (cw) and pulsed laser systems and the corresponding required laser intensities are calculated. In Section \ref{S:3} the cooling process is discussed and the required cooling time is estimated. Then, the influences of intrabeam scattering (IBS) and space charge (SC) effects on laser cooling are discussed and the maximum particle intensities are given in section \ref{S:4}. The paper is concluded with a comparison of two examples in section \ref{S:5} and an outlook in section \ref{S:6}.

\section{Laser Force}\label{S:2}

Laser cooling is based on the repetitive resonant absorption of unidirectional laser photons, followed by the spontaneous and directionally random emission of fluorescence photons. The momentum change of the ion during the interaction is a combination of the momenta of the incoming and outgoing photons. Each scattering event changes the momentum of the ion by (for more information see ref. \citep{metcalf1999})

\begin{align}
	\Delta p^{PF} = \Delta p^{PF}_{emit} - \Delta p^{PF}_{absorb} = - \frac{2\pi \hbar}{\lambda^{PF}}\cdot (1 + U_i),
\end{align}
where $\lambda^{PF}$ is the photon wavelength in the rest frame of the ions (PF), that has to match to the atomic transition of the ion. $U_i$ is a uniformly distributed random number in the interval $[-1,1]$ and describes the projection of the spontaneously emitted photon on the longitudinal axis of the accelerator. Applying the Lorentz transformation to the incoming and outgoing photons, the momentum kick in the laboratory frame (LF) results in

\begin{align}
	\Delta p^{LF} =\frac{2\pi \hbar}{\lambda^{LF; in}}\cdot \gamma^2 \cdot (1+\beta)\cdot (1+U_i), \label{equ:potonkick}
\end{align}
where $\lambda^{LF; in}$ describes the wavelength of the incoming laser photon in LF (see ref. \cite{Schramm2004-2}). The frequency of occurrence of a scattering event is given by the spontaneous emission rate $k_{se}^{PF}$, that is calculated by (see ref. \cite{metcalf1999})

\begin{align}
	k_{se}^{PF}(\delta,t) &= \rho_{ee}(\delta,t) \cdot \frac{1}{\tau_{se}^{PF}} \label{equ:scatrate}
\end{align}
where $\rho_{ee}(\delta,t)$ is the excitation probability (calculated in section \ref{S:2.1} and \ref{S:2.2}) and $\tau_{se}^{PF}$ the lifetime of the excited state. The integration of the emission rate over a time interval $\left[t_1,t_2\right]$ results in the average number of scattered photons per ion

\begin{align}
	n_{scat}(\delta) &= \int_{t_1}^{t_2}  k_{se}^{PF}(\delta,t) dt.
\end{align}
Neglecting the statistical component, the strength of the ion laser interaction can be expressed by an averaged force acting on the ions:

\begin{align}
\left<F_L^{LF}(\delta)\right>=\left<\Delta p^{LF}\right> \cdot \left<k_{se}^{LF}(t,\delta)\right> \label{equ:flaser}
\end{align}
The averaged momentum change during one turn of the ions in the accelerator, is given by

\begin{align}
	\Delta p_{turn}^{LF} (\delta) = \int_{0}^{\Delta t} \left<F_L^{LF}(\delta)\right> dt=\left<\Delta p^{LF}\right> \cdot n_{scat}(\delta). \label{equ:Favg}
\end{align}

\begin{figure}[ht]
	\centering
	\includegraphics[angle=-90, width=.47\textwidth]{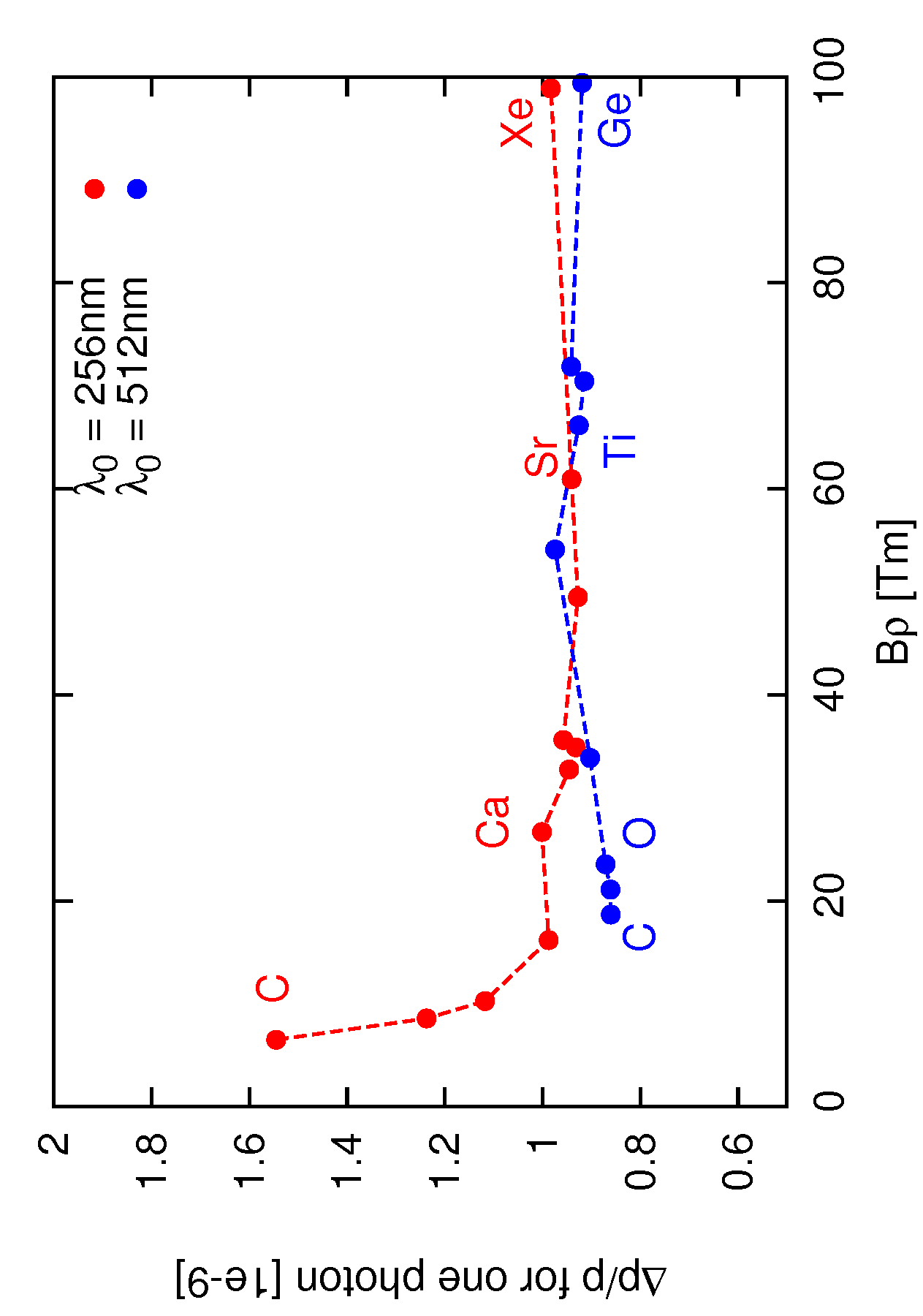}
	\caption{Mean value of relative momentum kick for one scattering event, calculated for the $2s_{1/2} \rightarrow 2p_{1/2}$ transition in Li-like ions. The transfered momentum stays nearly constant for different magnetic rigidities. Different dots represent different ions, which are partly marked by their element name.}
	\label{fig:singleKick}
\end{figure}

For a comparison of the laser force at different beam energies, the momentum kick is normalized to the initial momentum of the ion $p_0$.

\begin{align}
	\Delta \delta^{LF} = \frac{\Delta p^{LF}}{p_0}
\end{align}
For the $2s_{1/2}\rightarrow 2p_{1/2}$ transition in Li-like ions, the relative momentum kick of a single scattering event is very similar for different ions, as depicted in fig. \ref{fig:singleKick}. The values of the atomic transitions are taken from ref. \cite{johnson1996}. The dependency of the magnetic rigidity ($B\rho = p_0/q$) on the transition of the ion and the applied laser system is discussed in ref. \cite{schramm2004}. The relative momentum kick is  $\Delta \delta^{LF} \approx 10^{-9}$ and only increases slightly for very light ions, where $\beta < 1$. Therefore the impact on beam dynamics for a single scattering event is similar for ions at non-relativistic and relativistic beam energies. Transforming the momentum kicks of the absorbed (directional) and emitted (random) photons separately into the LF the ratio of these values is given by

\begin{align}
	\frac{\left<\Delta p^{LF}_{emit}\right>}{\Delta p^{LF}_{absorb}} = \gamma^2 (1+\beta)^2 - 1.
\end{align}
This equation shows that for non-relativistic beams the transferred momentum of the emitted photons can be neglected, whereas for relativistic beams the transferred momentum of the emitted photons dominates the scattering event.

For the calculation of the laser force, the population density of the excited state ($\rho_{ee}$ in eq. \ref{equ:scatrate}) is calculated by the optical Bloch equations (see ref. \cite{metcalf1999}). The equations can be solved for arbitrary laser intensities $I(t)$. In the following the excitation probabilities for two different laser scenarios are discussed. For simplicity we assume that the transverse laser beam spot covers the whole particle beam equally.

\subsection{Continuous Wave Laser}\label{S:2.1}
The ordinary solution for laser cooling experiments is to use a continuous wave (cw) laser system. The ions circulate in the accelerator and interact with the laser along a straight section with length $L_{interact}$. The ions see a rectangular laser pulse with length:

\begin{align}
	\Delta t_{cw}^{PF} = \frac{L_{interact}}{\gamma \beta c_0}
\end{align}
Usually the interaction section is long enough to approximate the excitation by a steady state solution ($\Delta t_{cw}^{PF} \gg \tau_{se}^{PF}$). For a constant excitation probability ($\dot\rho_{ee}=0$) the optical Bloch equations can be solved analytically \cite{metcalf1999} and the number of scattering events per turn is given by:

\begin{align}
	n_{scat}(\delta) &= \frac{L_{interact}}{\gamma \beta c_0} \cdot \frac{1}{2\tau_{se}^{PF}}\frac{S}{1+S+(2 \zeta (\delta-\delta_{LPos}) \cdot \tau_{se}^{PF})^2} \label{equ:nscat}\\
	\zeta &= \frac{d \omega}{d \delta} = \frac{2\pi c_0}{\lambda_{PF}}\beta \gamma (1+\beta)
\end{align}
where $S=\frac{I^{PF}}{I_s^{PF}}=\frac{I^{LF}}{I_s^{LF}}$ describes the saturation parameter, $\delta$ the relative momentum deviation of a test particle and $\delta_{LPos}$ the position of the laser in units of relative momentum. The width of the function $n_{scat}(\delta)$ and consequently the width of the laser force in units of relative momentum is given by:

\begin{align}
	\Delta_{fwhm} = \frac{\sqrt{1+S}\cdot \lambda^{LF}}{2\pi \tau^{PF}_{se} (1+\beta)\beta \gamma c_0} \label{equ:dfwhm}
\end{align}
Assuming a saturated transition, the width of the laser force is very similar for different Li-like ions, $\Delta_{fwhm}\approx 4\cdot 10^{-8}$ in units of relative momentum. Compared to the typical momentum spread of a heavy ion bunch, the laser force is very narrow. However, the width of the laser force does not change at relativistic beam energies and therefore similar cooling methods, that were already used successfully for moderately relativistic beams, can be applied to relativistic ion beams.

\begin{figure}[ht]
	\centering
	\includegraphics[angle=-90, width=.47\textwidth]{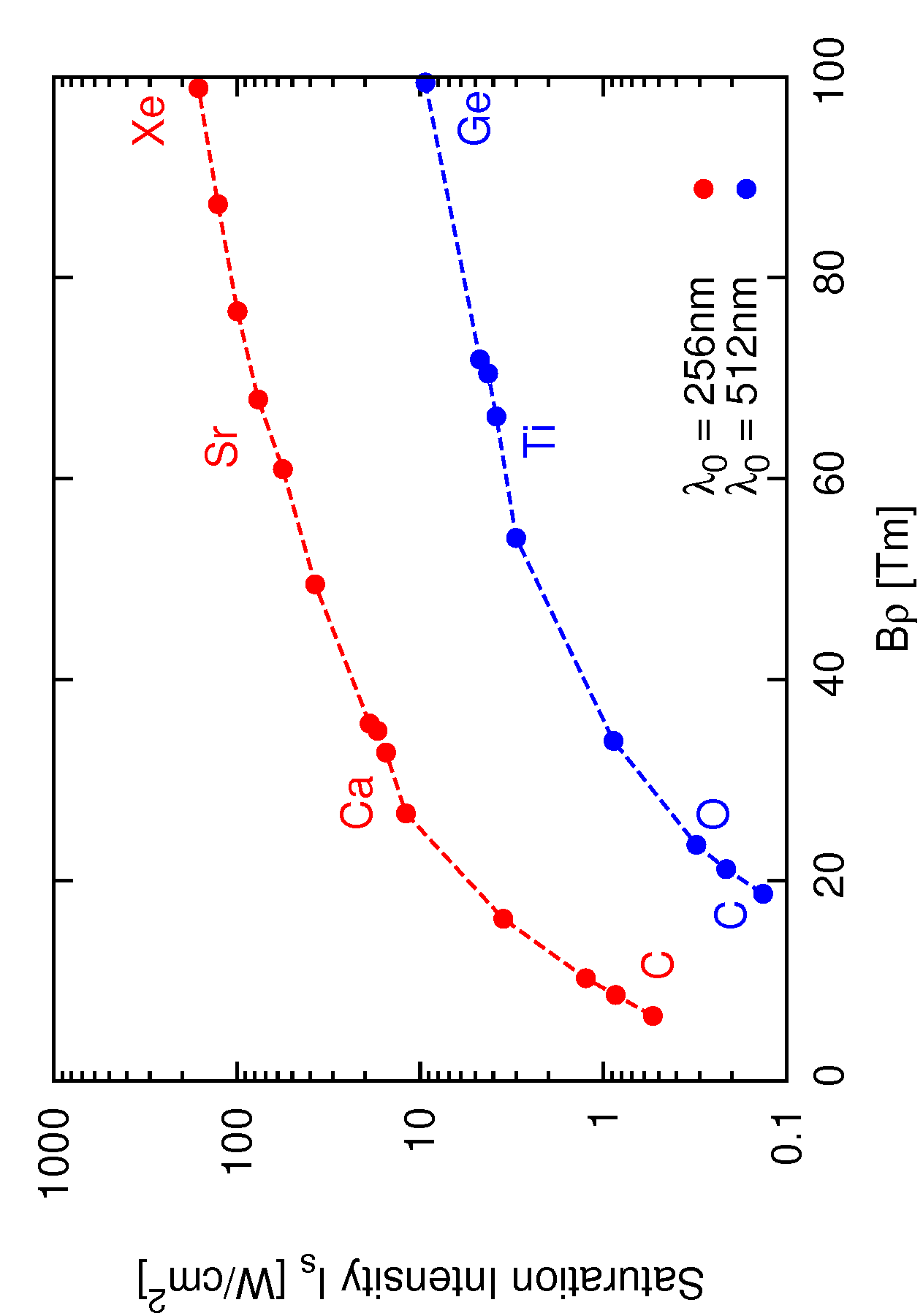}
	\caption{Calculated saturation intensities for different Li-like ions. For higher magnetic rigidities the required intensity increases significantly. Transition rates are given in ref. \cite{johnson1996}. Different dots represent different ions, which are partly marked by element name.}
	\label{fig:Isat}
\end{figure}

The required LF laser intensity for saturation is given by (see ref. \cite{Schramm2004-2})

\begin{align}
	I_s^{LF} &= I_s^{PF} \cdot \frac{1}{\gamma^2(1+\beta)^2} =  \frac{2\pi^2\hbar c_0}{3\lambda_{LF}^3 \tau^{PF}_{se}} \cdot \gamma(1+\beta). \label{equ:Isat}
\end{align}
The saturation intensities for the $2s_{1/2}\rightarrow 2p_{1/2}$ transition in Li-like ions for two different laser wavelengths are shown in fig. \ref{fig:Isat}. The saturation intensity increases strongly for higher beam energies. If the laser intensity is far below the saturation intensity of the transition, the number of scattering events decreases as described by eq. \ref{equ:nscat} and laser cooling becomes inefficient. At higher magnetic rigidities the required laser power for a typical beam cross-section of several $\mbox{mm}^2$ substantially exceeds the output power of existing UV laser systems ($256\,\mbox{nm}$). As an alternative laser cooling with a wavelength of $512\,\mbox{nm}$ looks promising, because the saturation intensity is lower and laser systems with significantly more laser power are available. More information about the laser system, which was successfully used in a previous laser cooling experiment (see ref. \cite{Winters2013}), is given in ref. \cite{Beck2016}.

\begin{figure}[ht]
	\centering
	\includegraphics[angle=-90, width=.47\textwidth]{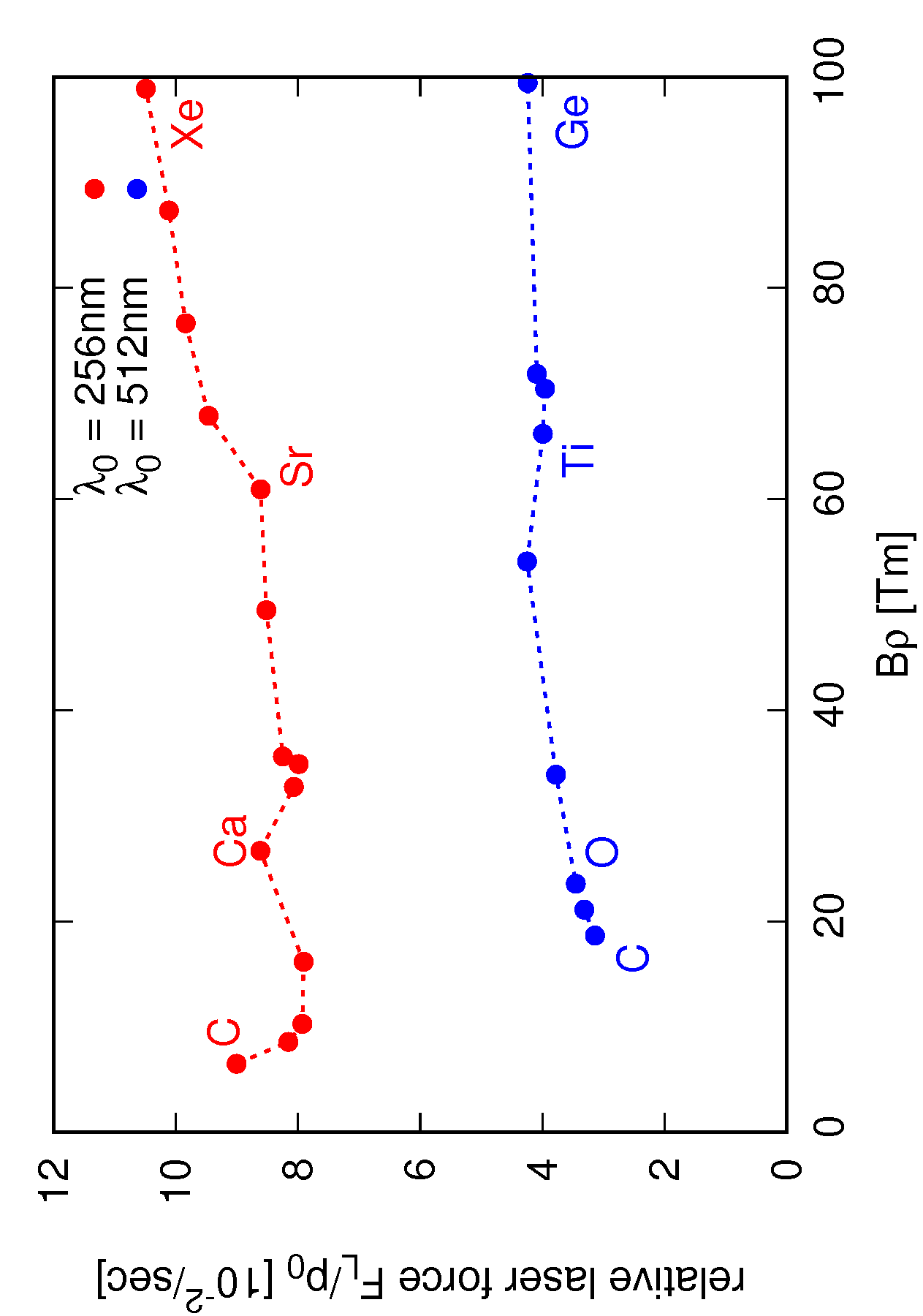}
	\caption{Laser force $F_L^{LF}$ divided by the initial momentum of the ions for different Li-like ions. Transition rates are given in ref. \cite{johnson1996}. Different dots represent different ions, which are partly marked by their element name. The transitions are assumed to be saturated.}
	\label{fig:flaser}
\end{figure}

A comparison of the relative laser force, i.e. eq. \ref{equ:flaser} divided by the initial momentum of the ion, for different Li-like ions is shown in fig \ref{fig:flaser}. The values are calculated for a saturated transition ($S=1$) and a continuous interaction of the laser beam with the particle beam. The relative force is equivalent to the relative momentum shift of an ion during one second, if the ion would be continuously in resonance with the laser during the time. The relative momentum kick per revolution of the particles in the accelerator can be evaluated by the multiplication of the relative force with the time, the ion needs to pass the interaction section. Assuming that enough laser power for saturation is available, the relative laser force of a cw laser system is very similar for non-relativistic and relativistic particle beams. This becomes clear from fig. \ref{fig:flaser}.

\subsection{Pulsed Laser}\label{S:2.2}
\begin{figure}[ht]
	\centering
	\includegraphics[angle=-90, width=.47\textwidth]{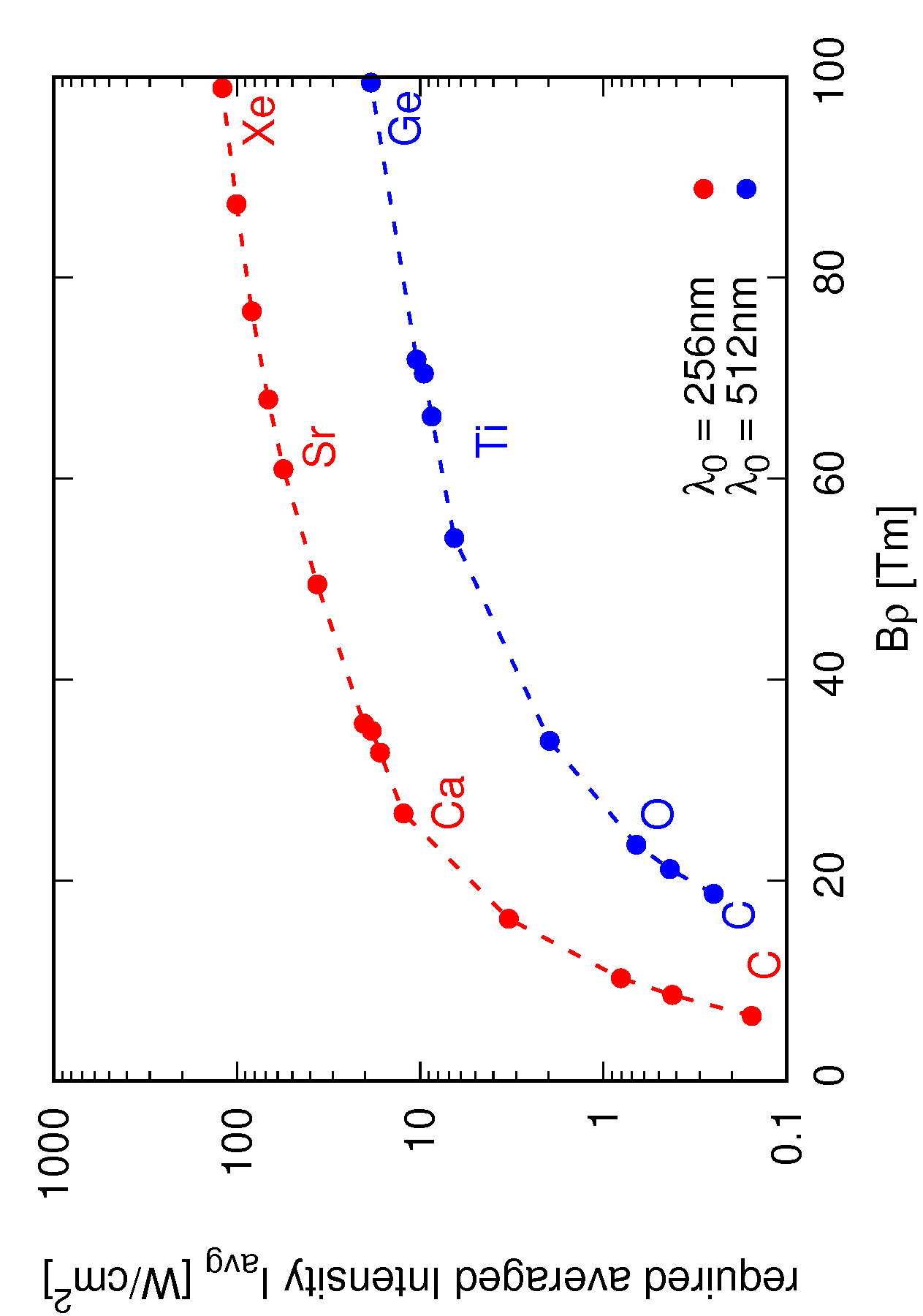}
	\caption{Required average intensity of pulsed laser system with $\sigma_{\delta}=10^{-5}$. The intensity scales linearly with $\sigma_\delta$. The values are calculated for $f_{rep} = f_{rev} \approx 278\,\mbox{kHz}$ in SIS100.}
	\label{fig:pulsedIavg}
\end{figure}

The second option is to use a pulsed laser system \cite{bussmann2014, Siebold2016}. Short laser pulses have, according to $\Delta f \cdot \Delta t \geq 0.5$, a correspondingly broad spectral width. For a Gaussian laser pulse with a standard deviation of $\sigma_{t}^{LF}$ in time, the width of the laser force, expressed in units of relative momentum, can be calculated by (see ref. \cite{bussmann2014}):

\begin{align}
	\sigma_\delta &= \frac{\sigma_p}{p}=\frac{1}{p}\frac{dp}{d\lambda^{LF}}\sigma_{\lambda}^{LF}= \frac{\lambda^{LF}}{2\pi c_0 \beta\sigma_{t}^{LF}}\\
	\sigma_{t}^{PF} &= \sqrt{\frac{1-\beta}{1+\beta}} \cdot \sigma_{t}^{LF}
\end{align}
In order to calculate the required laser intensity and the strength of the laser force, we assume that the pulses are short compared to the lifetime of the excited state ($ \sigma_{t}^{PF} \ll \tau_{se}^{PF}$). As a consequence, at most only one spontaneous emission per laser pulse can take place. 
The goal of the excitation with a short laser pulse is to have as many ions as possible in the excited state at the end of the pulse. After the pulse, every ion in the excited state spontaneously emits one photon. Solving the optical Bloch equations, the optimum average laser intensity is found to be

\begin{align}
I_{avg} = f_{rep} \cdot \sqrt{2\pi}\cdot \frac{\pi}{2}I_{S}^{LF}\cdot \frac{\tau_{se;\,PF}^2}{\sigma_{t}^{LF}}\cdot \frac{1+\beta}{1-\beta},
\end{align}
where all parameters are in the LF, except the lifetime of the excited state $\tau_{se}^{PF}$. The formula is derived in detail in \ref{A:2}. The required average intensity for the pulsed laser is depicted in fig. \ref{fig:pulsedIavg}. The intensity scales linearly with the spectral width, respectively inversely proportional with the pulse length.
The probability of a scattering event in the interaction section is given by:

\begin{align}
	\rho_{scat}(\delta) &= \rho_{excit}(I) \cdot \rho_{syn} \cdot e^{-\frac{1}{2}\left(\frac{\delta_{LPos}-\delta}{\sigma_{\delta}}\right)^2} \label{equ:rhoScatPulsed}\\
	\rho_{syn} &= \left\{\begin{array}{ll}
        \frac{f_{rep}}{f_{rev}}\cdot \frac{(1+\beta)\cdot L_{interact}}{L_{acc}}, & \text{(not synchronized) }\\
        \frac{f_{rep}}{f_{rev}} , & \text{(}f_{rep} < f_{rev} \text{ synchronized)}\\
        1 , & \text{(} f_{rep} \geq f_{rev} \text{ synchronized)}
        \end{array}\right.
\end{align}
where $\rho_{excit}(I)$ is the excitation probability $\rho_{ee}$ directly after the laser pulse (see eq. \ref{equ:rabiosc}). In case the repetition rate $f_{rep}$ is not synchronized to the revolution frequency of the bunches $f_{rev}$ the position of the laser pulses is randomly distributed and all bunches in the ring will interact with the laser. 

For a synchronized condition one bunch interacts with the laser pulses and can be cooled more efficiently. Note that for long bunches the laser acts only on ions in an area of $\left(1+\beta\right) L_{interact}$. On a small accelerator like ESR the synchronization only results in a gain of 2.9 whereas for SIS100 the factor is 20.8.

The averaged relative momentum kick per turn of a pulsed laser results in

\begin{align}
\Delta \delta_{turn}^{pulsed}(\delta) = \left<\Delta\delta^{LF}\right>\cdot \rho_{excit}(I)\cdot \rho_{syn}\cdot e^{-\frac{1}{2}\left(\frac{\delta_{LPos}-\delta}{\sigma_{\delta}}\right)^2}.
\end{align}
Comparing the strength of the pulsed laser system for non-relativistic and relativistic particle beams, the relative momentum kick per turn for a maximized $\rho_{scat}$ is $\Delta \delta_{turn}^{pulsed}(\delta_{LPos}) \approx 10^{-9}$ and very similar for different Li-like ions.

\section{Cooling Process}
\label{S:3}
\subsection{CW Laser} \label{S:3.1}

\begin{figure}[ht]
    \centering
    \includegraphics[width=40mm]{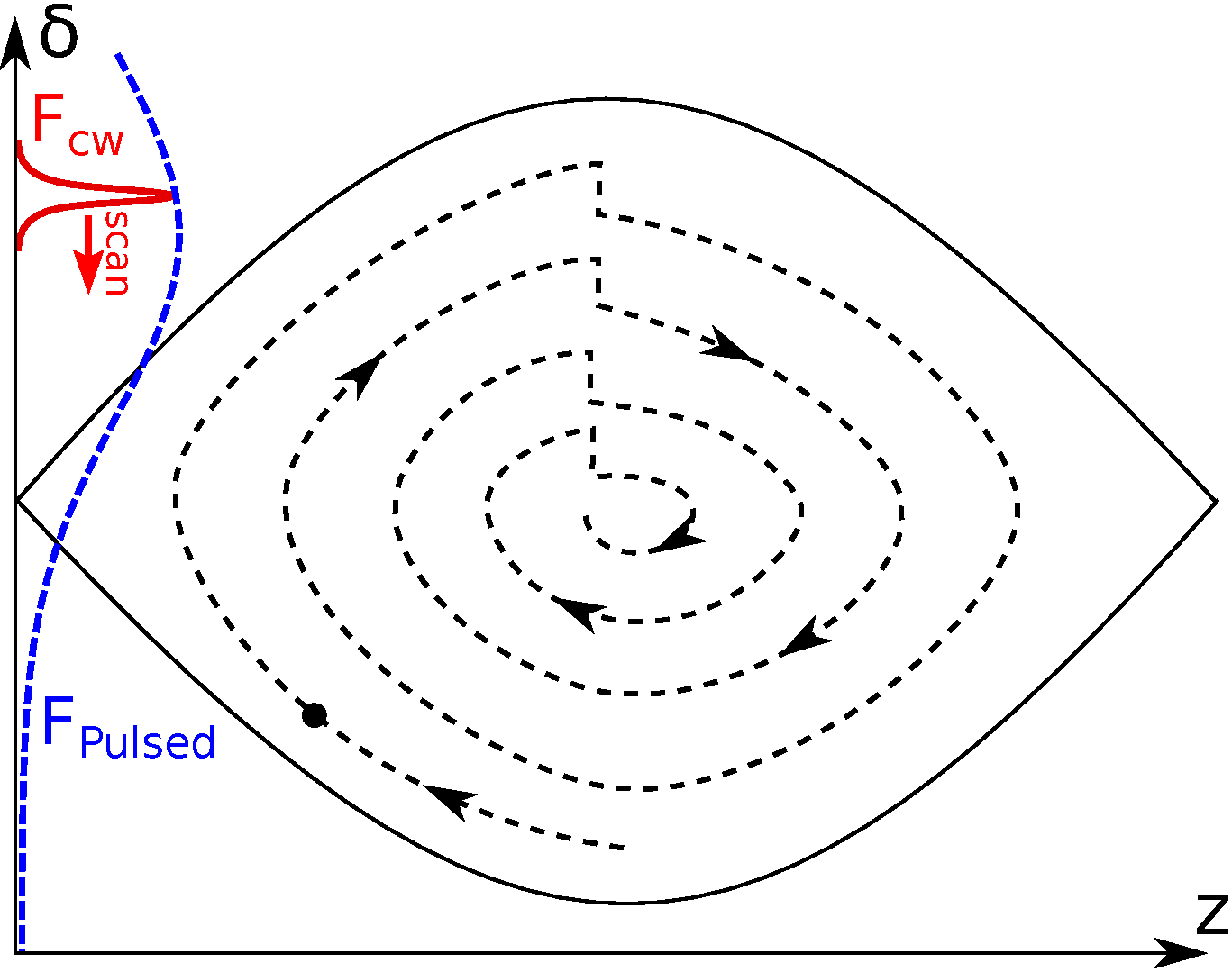}
    \caption{Sketch of the longitudinal phase space during the cooling process of a hot ion beam in a rf bucket. The laser force of a cw laser and the pulsed laser are indicated by the red and the dashed blue curve. The cw laser force is scanned to damp continuously the synchrotron oscillations of all ions whereas the pulsed laser force interacts with all ions without a scan.}
    \label{fig:bucket}
\end{figure}

The cooling process of a particle distribution requires a stable point, where the sum of all forces cancels. In order to achieve a stable point for the laser cooling process the laser force is counteracted by the rf voltage. For the initial hot ion bunch, the narrow laser force (see eq. \ref{equ:dfwhm}) does affect only a small fraction of the particles in phase space. Therefore the position of the laser force is scanned from the position of the particle with the highest oscillation amplitude to the synchronous particle in order to damp the oscillations of all ions in the bucket like shown in fig. \ref{fig:bucket}.

\begin{figure}[ht]
\centering
\includegraphics[angle=-90,width=.47\textwidth]{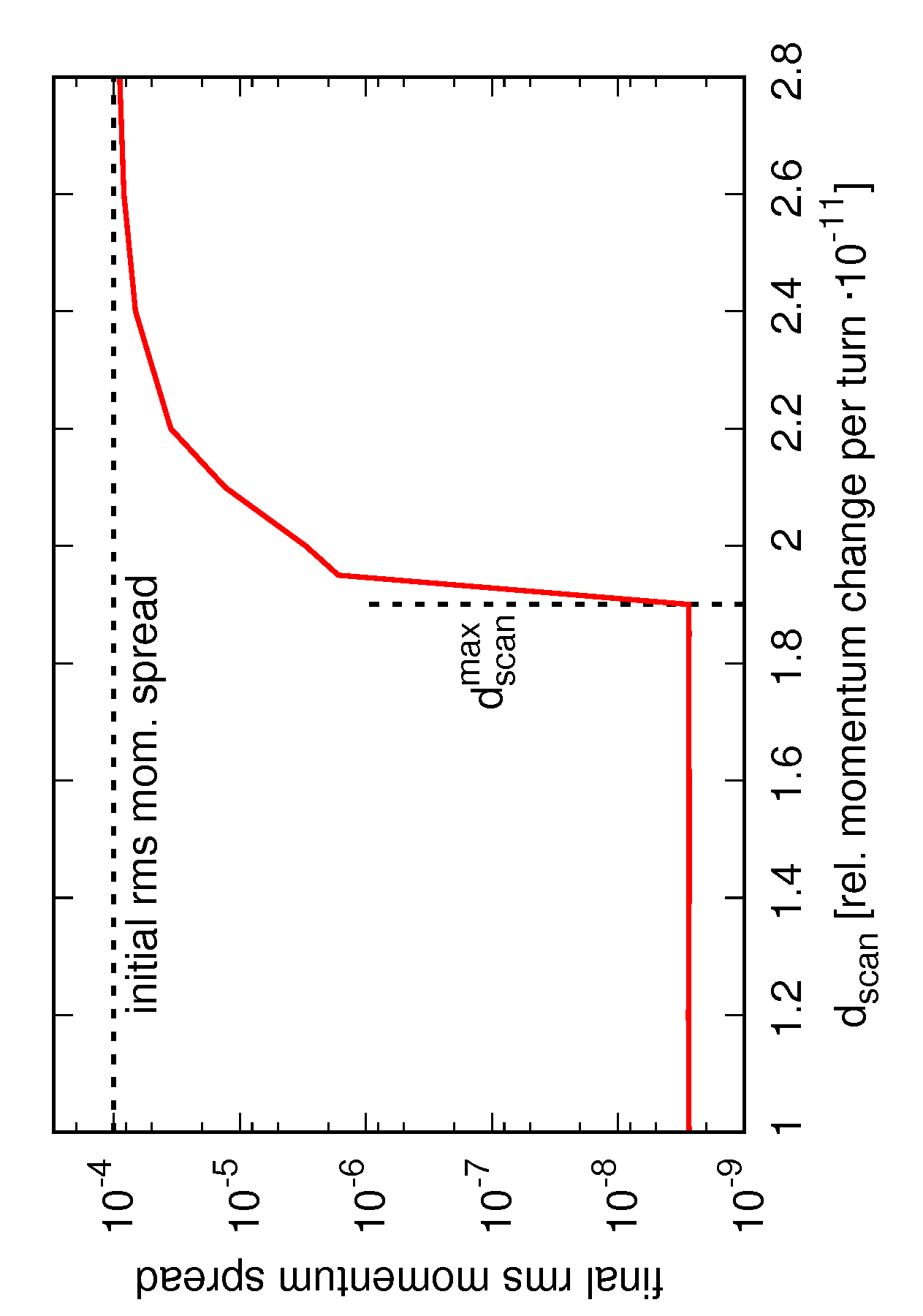}
\caption{Final rms momentum spread for different scan speeds of the laser force in units of relative momentum per turn for $\mbox{Ti}^{19+}$ ions in SIS100.}
\label{fig:scanLaser}
\end{figure}

The success and efficiency of the cooling process depends strongly on the speed of the laser force scan in phase space. The final rms momentum spread after the laser scan for different scan speeds $d_{scan}$ is shown in fig. \ref{fig:scanLaser}. The unit of the scan speed $d_{scan}$ is given by the change in relative momentum of the laser resonance per turn. The results are shown for the cooling process of $\mbox{Ti}^{19+}$ ions in SIS100. The detailed parameter list is given in tab. \ref{tab:ESRSIS} and \ref{tab:ESRSISLaser} in section \ref{S:5} except the bunch length that is set in this case to $4\,m$ in order to stay in the linear region of the bucket. The results in fig. \ref{fig:scanLaser} indicate, that below a certain speed of the laser scan (marked as $d_{scan}^{max}$) the cooling process is successful, whereas above this scan speed the rms momentum spread of the ion distribution stays close to the initial value.

\begin{figure}[ht]
\centering
\includegraphics[width=.47\textwidth]{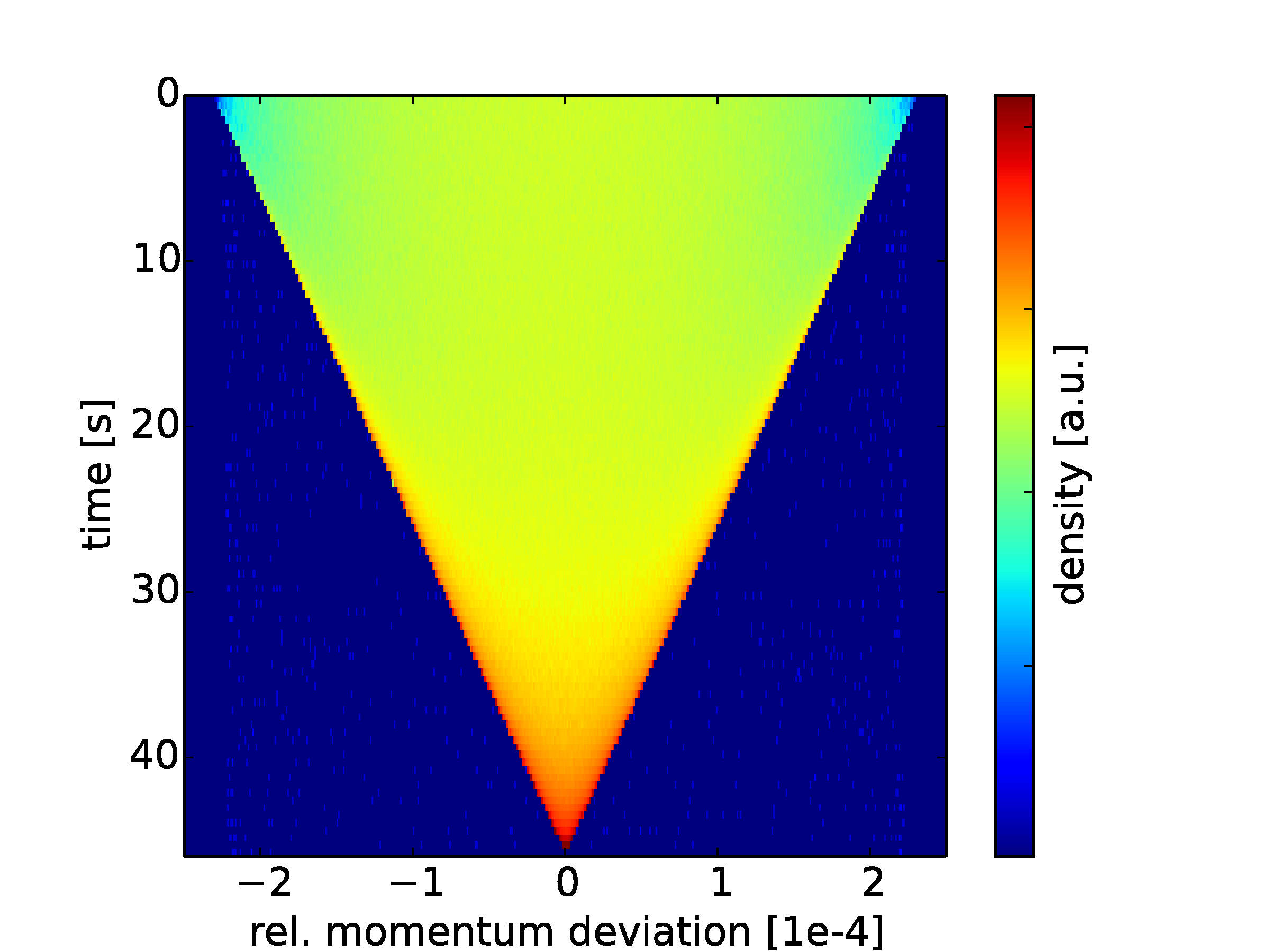}
\includegraphics[width=.47\textwidth]{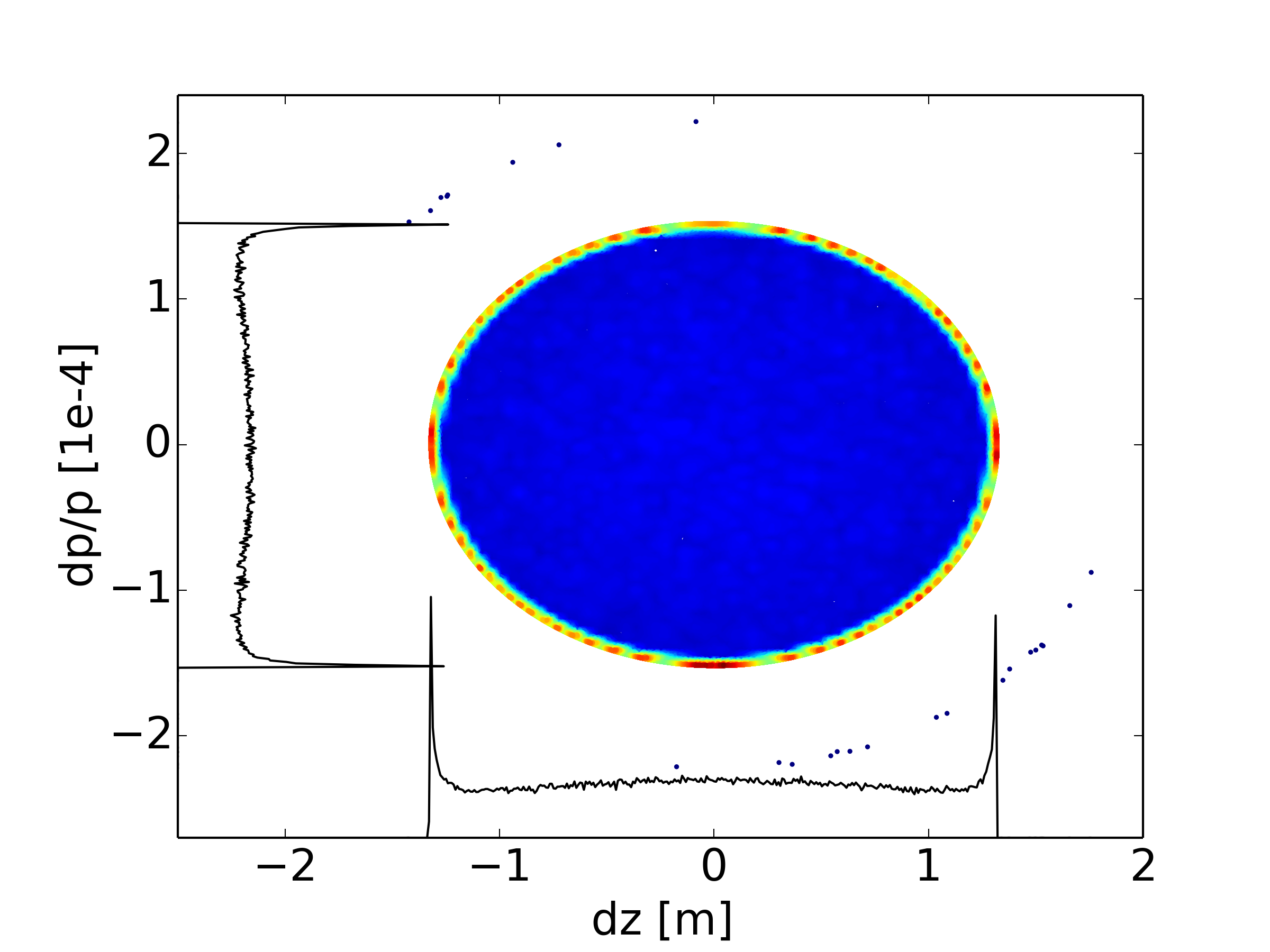}
\caption{Evolution of the momentum distribution over time during the cooling process and snapshot of the phase space distribution for a successful cooling process ($d_{scan}<d_{scan}^{max}$) of $\mbox{Ti}^{19+}$ ions in SIS100. The ions are pushed continuously in front of the laser force to the center of the bucket.}
\label{fig:scanSuccess}
\end{figure}

\begin{figure}[ht]
\centering
\includegraphics[width=.47\textwidth]{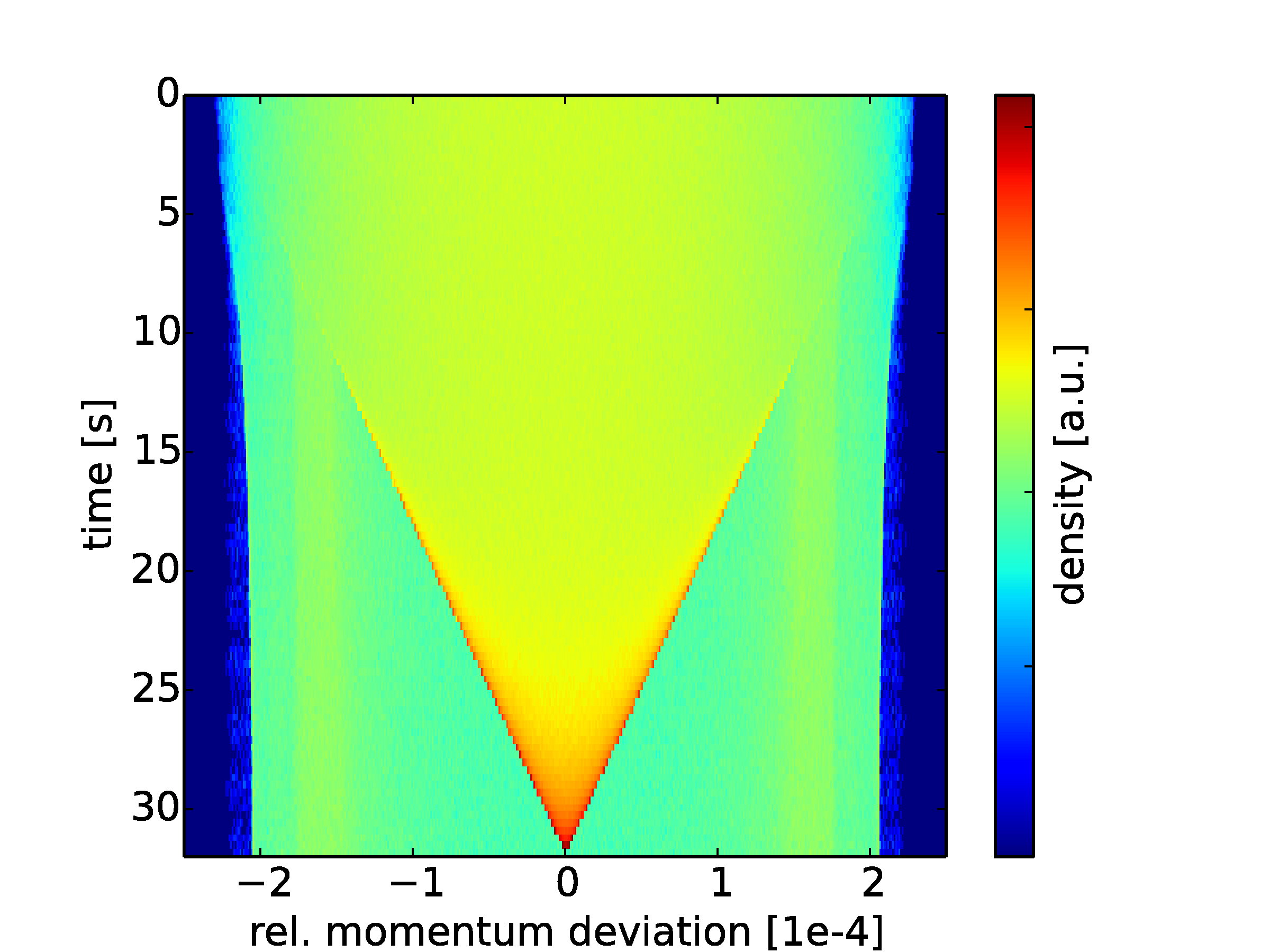}
\includegraphics[width=.47\textwidth]{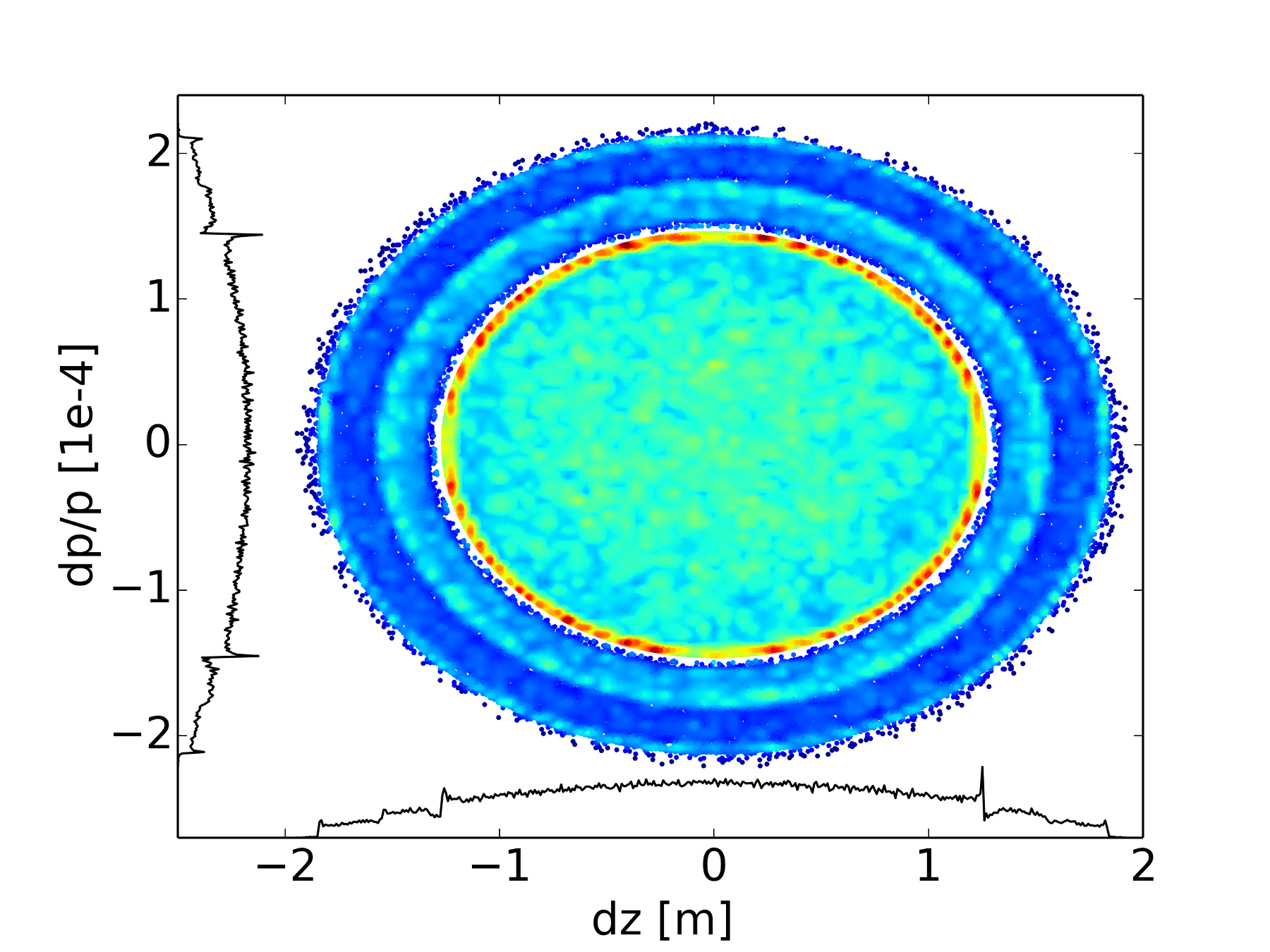}
\caption{Evolution of the momentum distribution over time during the cooling process and snapshot of the phase space distribution for an unsuccessful cooling process ($d_{scan}>d_{scan}^{max}$) of $\mbox{Ti}^{19+}$ ions in SIS100. The ions are lost behind the laser force and are no longer affected by the laser force.}
\label{fig:scanFailure}
\end{figure}

\begin{figure}[ht]
	\centering
	\includegraphics[width=.4\textwidth]{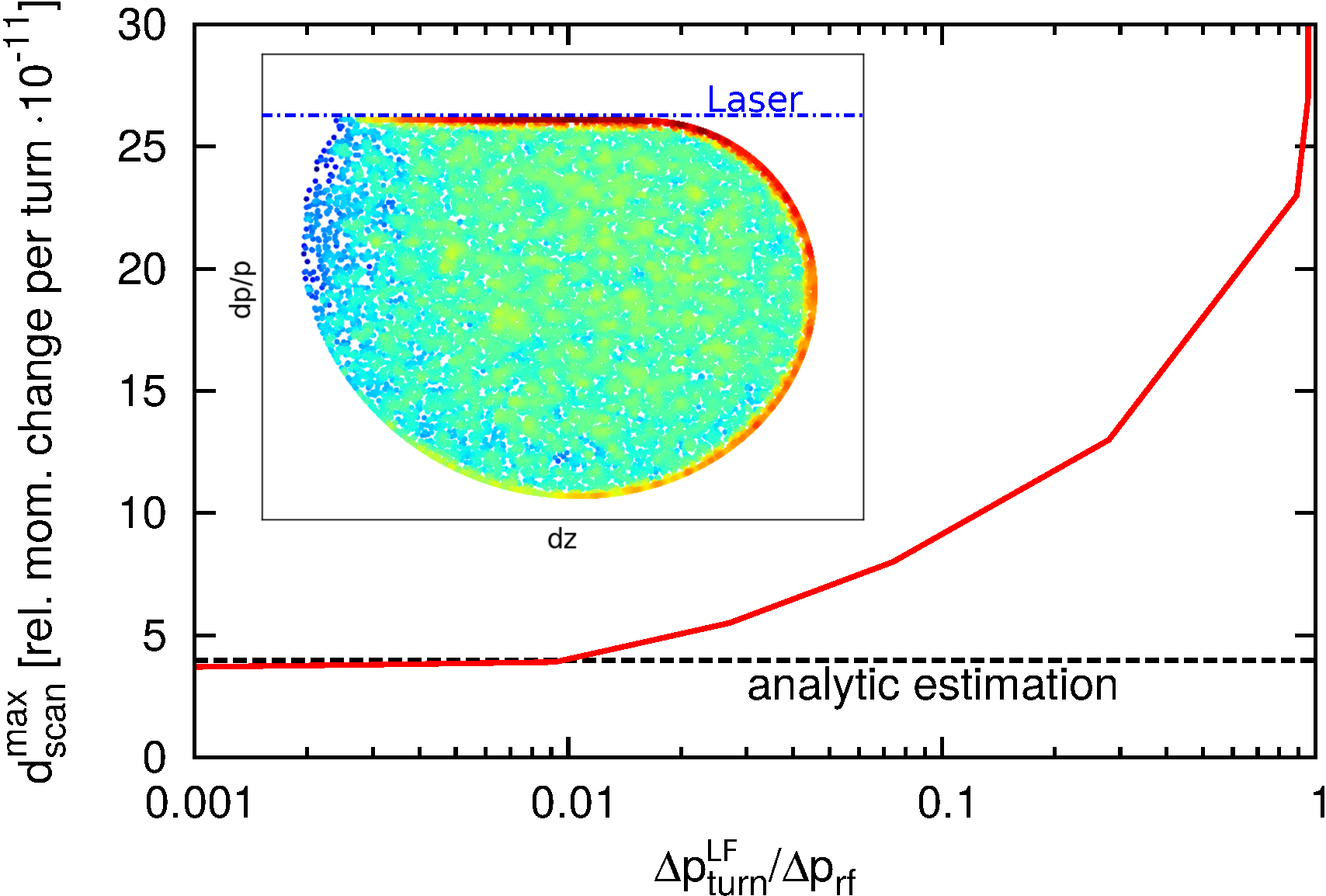}
	\caption{Maximum scan speed of the laser resonance in units of relative momentum per turn for different ratios of the momentum change of the laser force divided by the momentum change of the rf potential. The snapshot of the phase space density for $\Delta p_{turn}^{LF}/\Delta p_{rf} \approx 1$ shows that the ions stay in resonance with the laser light for a long time, which increases the maximum scan speed respectively reduces the required cooling time.}
	\label{fig:scanSpeeds}
\end{figure}

The evolution of the momentum distribution and a snapshot of the phase space distribution is shown in fig. \ref{fig:scanSuccess} for a successful cooling process ($d_{scan} < d_{scan}^{max}$) and in fig. \ref{fig:scanFailure} for an unsuccessful cooling process ($d_{scan} > d_{scan}^{max}$). The cooling process is successful if all ions stay always in front of the laser force and move with repetitive steps to the center of the bucket. For $d_{scan} > d_{scan}^{max}$ the laser force moves faster to the center of the bucket than the ions. Consequently the ions are left behind the laser force and do not interact with the laser light for the rest of the cooling process. Therefore the maximum laser scan speed $d_{scan}^{max}$ is equal to the momentum change of the ions per turn averaged over the synchrotron motion.

\begin{align}
	d_{scan}^{max}\approx \left<\Delta\delta^{cw}_{turn}\right>_{syn} \label{equ:dscan}
\end{align}
The averaged momentum change is calculated for the ions with the oscillation amplitude equal to the resonant momentum of the laser force ($\hat \delta = \delta_{LPos}$). The laser affects the ions only on a small fraction of the circular motion (see fig. \ref{fig:bucket}). The average of the laser kick $\Delta\delta_{turn}^{LF}(\delta)$ over one oscillation period is calculated by:

\begin{align}
	\left<\Delta\delta^{cw}_{turn}\right>_{syn} = \frac{1}{2\pi}\int^{2\pi}_0 \Delta\delta_{turn}^{LF}\left(\delta_{LPos}\cdot \cos(\phi) \right)\cdot \cos(\phi) d\phi \label{equ:dscanmax}
\end{align}
This estimation is only valid for a small perturbation of the circular motion by the laser force, which is typically the case for high synchrotron tunes ($Q_s \approx 10^{-3}$). At higher energies and correspondingly lower synchrotron tunes $\left<\Delta\delta_{turn}\right>_{syn}$ can differ. The strength of the perturbation is expressed by the ratio of the momentum change of the laser force (see eq. \ref{equ:Favg}) divided by the momentum change of the rf potential $\Delta p_{turn}^{LF}/\Delta p_{rf}$. Figure \ref{fig:scanSpeeds} shows the maximum scan speed for different strengths of the perturbation. The results are given for the cooling process of $\mbox{Ti}^{19+}$ ions in SIS100 (except $\Delta \delta^{LF}$ is multiplied by two in order to cover the whole range). The ratio of the laser and rf force is changed by varying the strength of the rf potential respectively the bunch length. If $\Delta p_{turn}^{LF}/\Delta p_{rf} \ll 1$ (small perturbation) the simulation results converge towards the analytic estimation (see eq. \ref{equ:dscanmax}) while for $\Delta p_{turn}^{LF}/\Delta p_{rf} \approx 1$ (strong perturbation) the maximum laser scan speed increases. The strong perturbation of the circular motion in the bucket, like illustrated in the phase space plot in fig. \ref{fig:scanSpeeds}, keeps the ions in resonance with the laser light for a longer time and increases the possible scan of the laser resonance per turn.

Due to the linear scan of the laser force from the particle with the highest oscillation amplitude $\hat \delta_m$ to the center of the bucket, the required cooling time is given by:

\begin{align}
	T_{cool} = T_{rev} \cdot \frac{\hat\delta_m}{d_{scan}} \label{equ:Tcoolcw}
\end{align}
Because the duration of the cooling process is inversely proportional to the scan speed, the fastest successful cooling process can be achieved by choosing the scan speed $d_{scan}^{max}$. On the other hand this value can be reduced by reducing the ratio $\Delta p_{turn}^{LF}/\Delta p_{rf}$. Therefore low synchrotron tunes are beneficial for a fast cooling process.

\begin{figure}[ht]
\centering
\includegraphics[width=.4\textwidth]{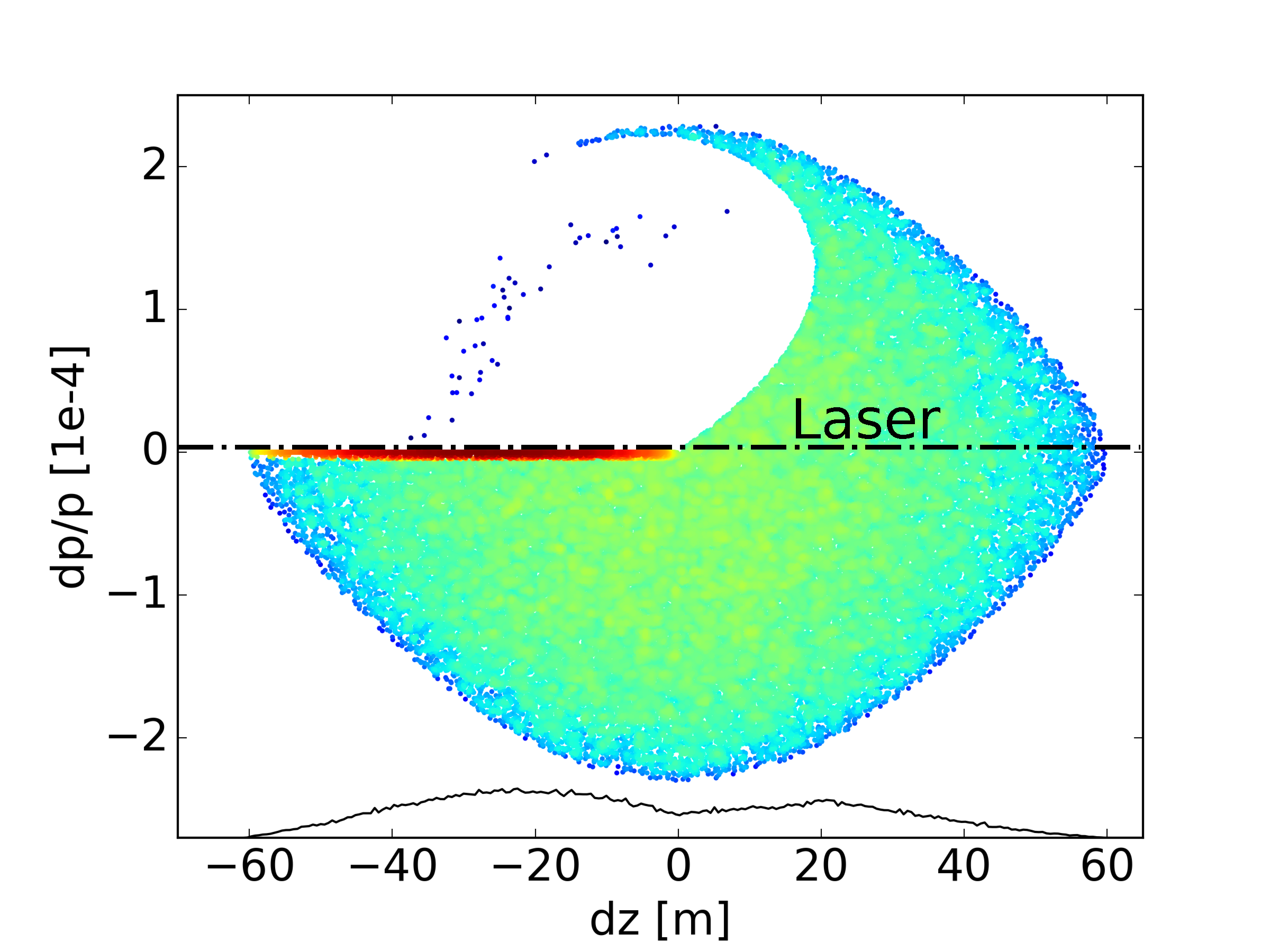}
\caption{Phase space plot of an alternative cooling scheme for a strong laser force $\Delta p_{turn}^{LF}/\Delta p_{rf} > 1$. The laser resonance is set close to the center of the bucket and the ions rotate into the laser force. At the position of the laser force the synchrotron motion of the ions is stopped.}
\label{fig:fixedLaserScheme}
\end{figure}

An alternative cooling scheme exists if the laser force is stronger than the maximum rf force $\Delta p_{turn}^{LF}/\Delta p_{rf} > 1$. If this condition is fulfilled the laser resonance can directly be set close to the center of the bucket. The particles are not able to pass the laser force because the laser is stronger than the rf force. As a result the synchrotron motion of the ions is interrupted and the particles will stay close to the laser force as illustrated in fig. \ref{fig:fixedLaserScheme}. The ions, that are not in resonance with the laser, perform the synchrotron motion until they are captured by the laser force. Therefore the cooling time is reduced to one synchrotron period

\begin{align}
	T_{cool}^{fixedCW}=T_{rev}\cdot \frac{1}{Q_s}
\end{align}
where $Q_s$ is the synchrotron tune. This cooling process is much faster compared to the scan of the laser resonance, but this scheme is only possible for a strong laser cooling force and a slow motion of the ions in the bucket. If the condition for this cooling scheme is fulfilled ($\Delta p_{turn}^{LF}/\Delta p_{rf} > 1$), a further reduction of the synchrotron tune increases the cooling time.

\subsection{Pulsed Laser}
\begin{figure}[ht]
	\centering
	\includegraphics[angle=-90, width=.47\textwidth]{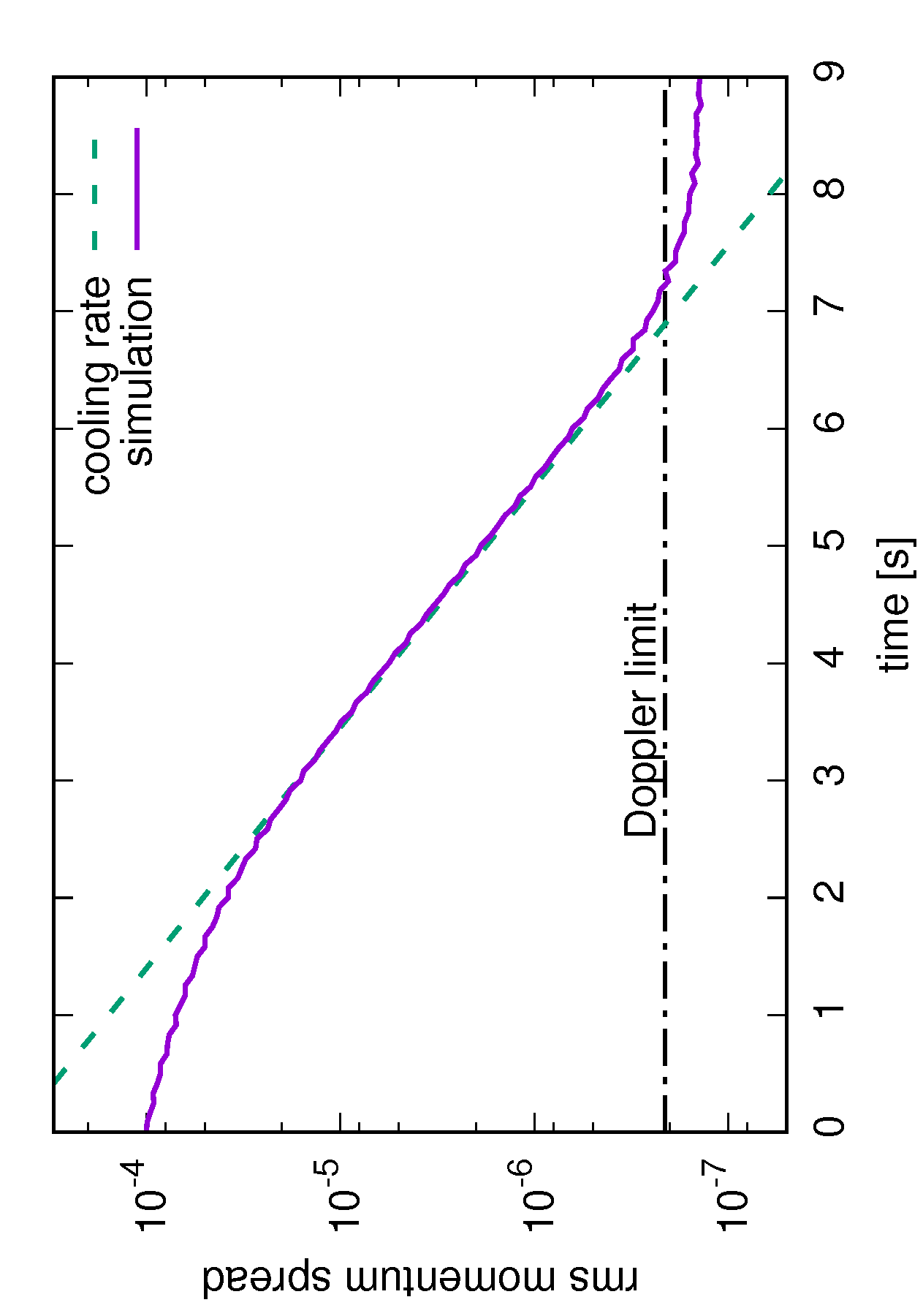}
	\caption{Cooling process of $\mbox{Ti}^{19+}$ ions in SIS100 with a single pulsed laser system. The dashed lines show the analytically calculated cooling rate (eq. \ref{equ:pulseCoolRate}) and the Doppler limit (eq. \ref{equ:pulsedLimit}).}
	\label{fig:CoolPulsed}
\end{figure}
In contrast to the narrow band cw laser force the pulsed laser affects all ions in phase space. Therefore no frequency scan is required (see fig. \ref{fig:bucket}). In combination with the rf force a stable point is created in the center of the bucket that is surrounded by an approximately linear cooling force. The result is an exponential decreasing rms momentum spread with the cooling rate:

\begin{align}
	\tau^{-1}_{cool} &= \frac{1}{T_{rev}}\frac{\partial\left<\Delta\delta_{turn}^{pulsed}\right>_{syn}}{\partial \delta}\label{equ:pulseCoolRate}\\
	&\approx-\frac{1}{T_{rev}}\cdot \left<\Delta \delta^{LF}\right> \cdot \rho_{scat}(0) \frac{\delta_{LPos}}{2\sigma_\delta^2} \label{equ:pulseFcool}
\end{align}
where $\left<\Delta\delta_{turn}^{pulsed}\right>_{syn}$ is the momentum kick of the pulsed laser averaged over a synchrotron period as calculated by eq. \ref{equ:dscanmax} for the cw laser system. As shown in fig. \ref{fig:CoolPulsed} the linear approximation of the cooling force describes the simulated cooling process of $\mbox{Ti}^{19+}$ ions in SIS100 properly. The simulation model is described in \ref{A:1} and the detailed parameter list is given in tab. \ref{tab:ESRSIS} and \ref{tab:ESRSISLaser} in section \ref{S:5} except the excitation probability $\rho^{pulsed}_{excit} = 1$.

In contrast to the cooling process with the cw laser system, the distribution stays close to a Gaussian shape during the whole cooling process because of the linear cooling force. The required time for the cooling process is given by

\begin{align}
	T_{cool} = \tau_{cool} \cdot \ln\left(\frac{\delta_0}{\delta_{f}}\right)
\end{align}
where $\delta_0$ and $\delta_{f}$ are the initial and final rms momentum spread of the bunch. The cooling is counteracted by the diffusive heating of the pulsed laser. The average momentum change of a scattering event is $\left<\Delta \delta^{LF}\right>$ and the average time between two scattering events is $T_{rev}/\rho_{scat}(0)$. The diffusion coefficient result in

\begin{align}
	D_{L}&=\frac{\left<\Delta \delta^{LF}\right>^2 \cdot \rho_{scat}(0)}{2T_{rev}}.
\end{align}
The lowest attainable rms momentum spread $\delta_{equ}^{pulsed}$ is given by the equilibrium state ($\tau_{cool}^{-1} + \tau^{-1}_{Lheat} = 0$) of the diffusive heating of the laser ion interaction and the cooling rate.

\begin{align}
	\delta_{equ}^{pulsed} &= \sqrt{\frac{\left<\Delta \delta^{LF}\right>\cdot \sigma_{\delta}^2}{\delta_{LPos}}} \label{equ:pulsedLimit}
\end{align}
This state is called Doppler limit (see ref. \cite{metcalf1999}) and is much higher for a pulsed laser system compared to a cw laser system. 

\section{Intensity Effects}
\label{S:4}

The final state of the cooling process will depend on intensity effects like intrabeam scattering (IBS) and space charge (SC). The heating rate of IBS increases for lower rms momentum spreads and directly counteracts the cooling process, whereas SC does not in general heat the beam but influences the trajectory of the ions in phase space. The SC potential for a bunch in an rf bucket is described by

\begin{align}
	V_{sc} &= -\frac{1}{2\pi}q\beta c L_{acc} X_{sc} \frac{\partial \lambda}{\partial z}\\
	X_{sc} &= \frac{g_{sc}}{2\epsilon_0 \beta c \gamma^2}
\end{align}
where $\lambda$ is the line density of the ion bunch and $g_{sc}$ the space charge g-factor (see ref. \cite{Boine2006,khateeb2001}). For an elliptical bunch distribution the SC potential is proportional to the applied rf potential. During the cooling process with a cw laser as well as a pulsed laser the particle density increases until the SC potential becomes equal to the applied rf potential ($V_{rf} = V_{sc}$). The bunch length $L_B$ of the equilibrium state is calculated by

\begin{align}
    L_{equ}=2\cdot \hat z_{equ} = \sqrt[3]{\frac{3 q L^2_{acc} g_{sc} N_p}{2\pi^2 h \epsilon_0\gamma^2 \hat V_{rf}}}. \label{equ:zMin}
\end{align}
The scaling of the bunch length with respect to the rf voltage and the particle number agrees with the experimental results presented in ref. \cite{Hangst1995}.

\begin{figure}[ht]
\centering
\includegraphics[width=.5\textwidth]{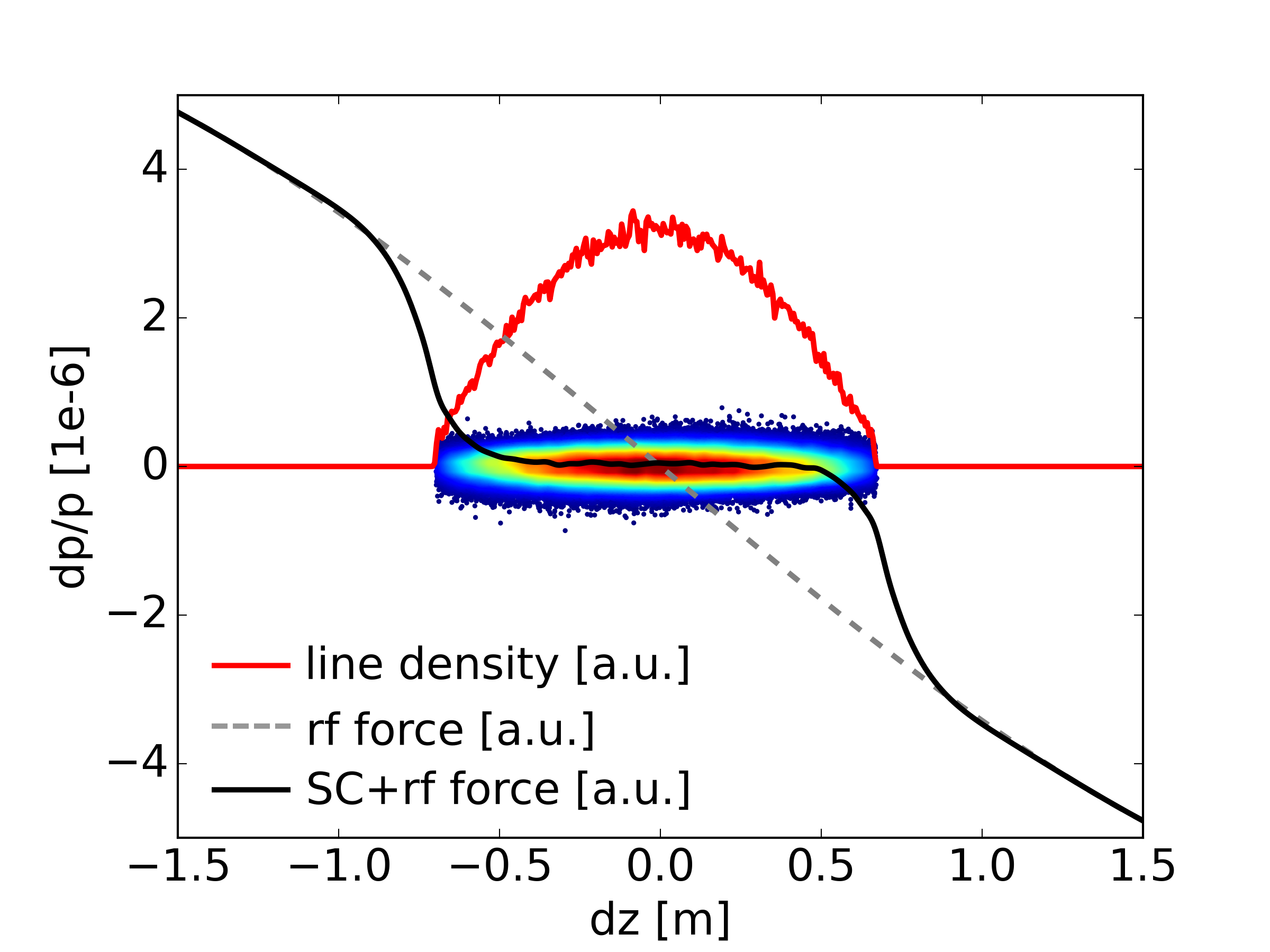}
\caption{Equilibrium state of SC and rf potential $V_{sc}=V_{rf}$. Inside the ion distribution the sum of the rf and SC force vanishes.}
\label{fig:zMinPhaseSpace}
\end{figure}

\begin{figure}[ht]
\centering
\includegraphics[angle=-90, width=.5\textwidth]{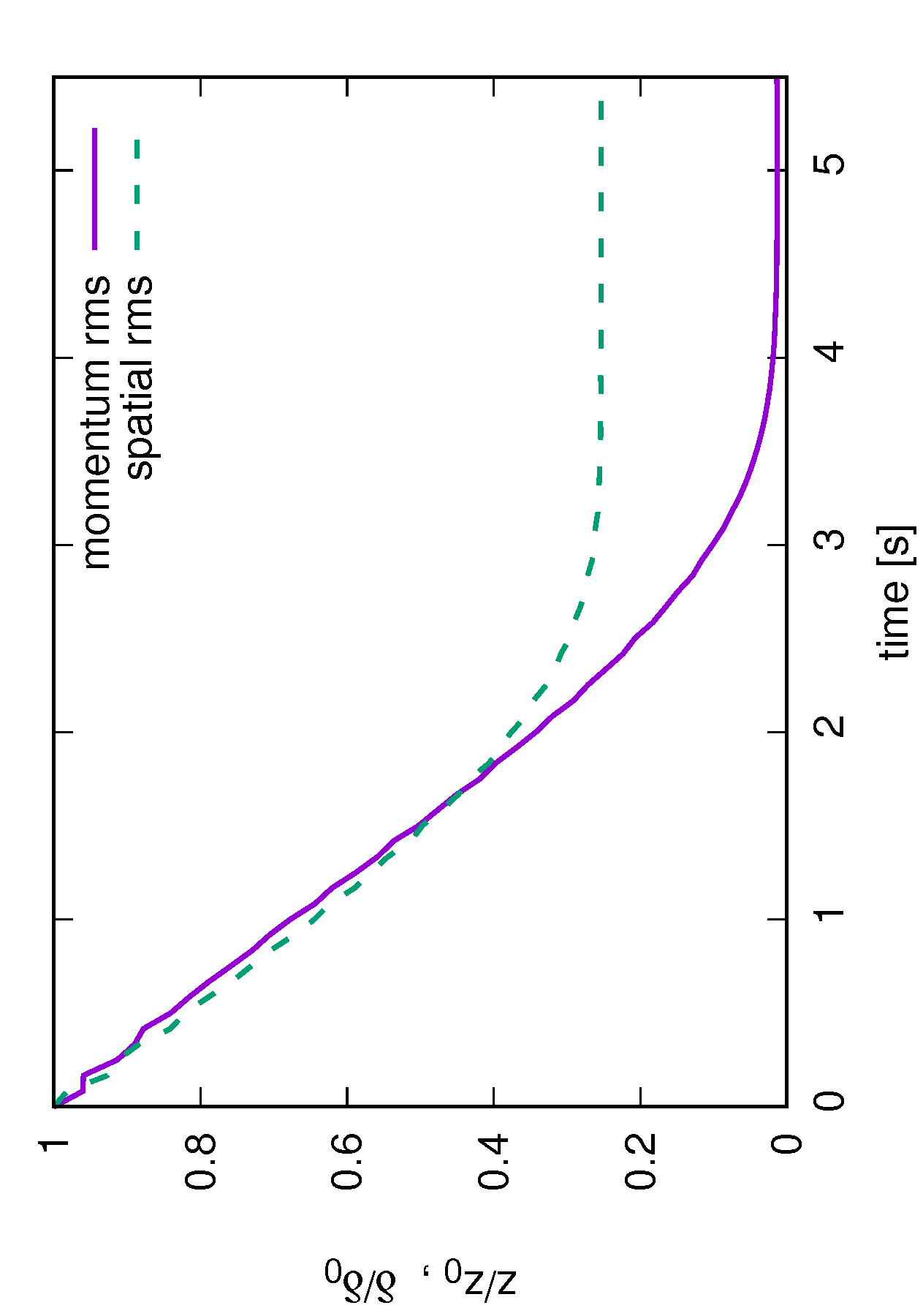}
\caption{Rms spatial and momentum spread, normalized to the initial rms value, during the cooling process with a pulsed laser for $\mbox{Ti}^{19+}$ ions in SIS100. Without SC the relative reduction of the spatial and momentum rms are equal, while in presence of SC the bunch length is only reduced until the equilibrium state is reached. The final bunch length can be calculated by eq. \ref{equ:zMin}.}
\label{fig:zMinRms}
\end{figure}

The phase space distribution and the sum of the rf and SC force is illustrated in fig. \ref{fig:zMinPhaseSpace}. The complete compensation of the rf potential inside the bunch prevents a further reduction of the bunch length. By increasing or decreasing the rf voltage the bunch length can be modified. However the equilibrium state of the SC and rf potential do not limit the rms momentum spread, as exemplary shown in fig. \ref{fig:zMinRms} for the cooling process with a pulsed laser. In absence of SC the reduction of the spatial and the rms momentum spread during the cooling process, normalized to the initial values, are correlated and after the cooling process both rms values are reduced by the same fraction. While in presence of SC the bunch length can only reach the equilibrium state whereas the rms momentum spread is reduced beyond the spatial rms spread and is limited by heating effects that are described in the following.

The effect of IBS can be described mathematically by the diffusion coefficient in the Boltzman-Fokker-Planck equation (see ref. \cite{Hannes1984}). The diffusion coefficient can be calculated by e.g. Bjorken-Mtingwa formula (see ref. \cite{Bjorken1982}). Assuming a much higher mean velocity deviation of the ions in transverse than in longitudinal direction, the major IBS diffusion coefficient $D_{IBS} = D_{zz}$ transfers heat from the transverse plain to the longitudinal coordinate. For higher relativistic factors the strength of IBS and SC decreases, in contrast the charge state of ions, cooled at higher relativistic factors, increases. As a result the IBS rate and SC strength is very similar for different Li-like ions, assuming equal bunch parameters.

\subsection{IBS effects for cw Laser}\label{S:4.1}

\begin{figure}[ht]
	\centering
	\includegraphics[angle=-90, width=.47\textwidth]{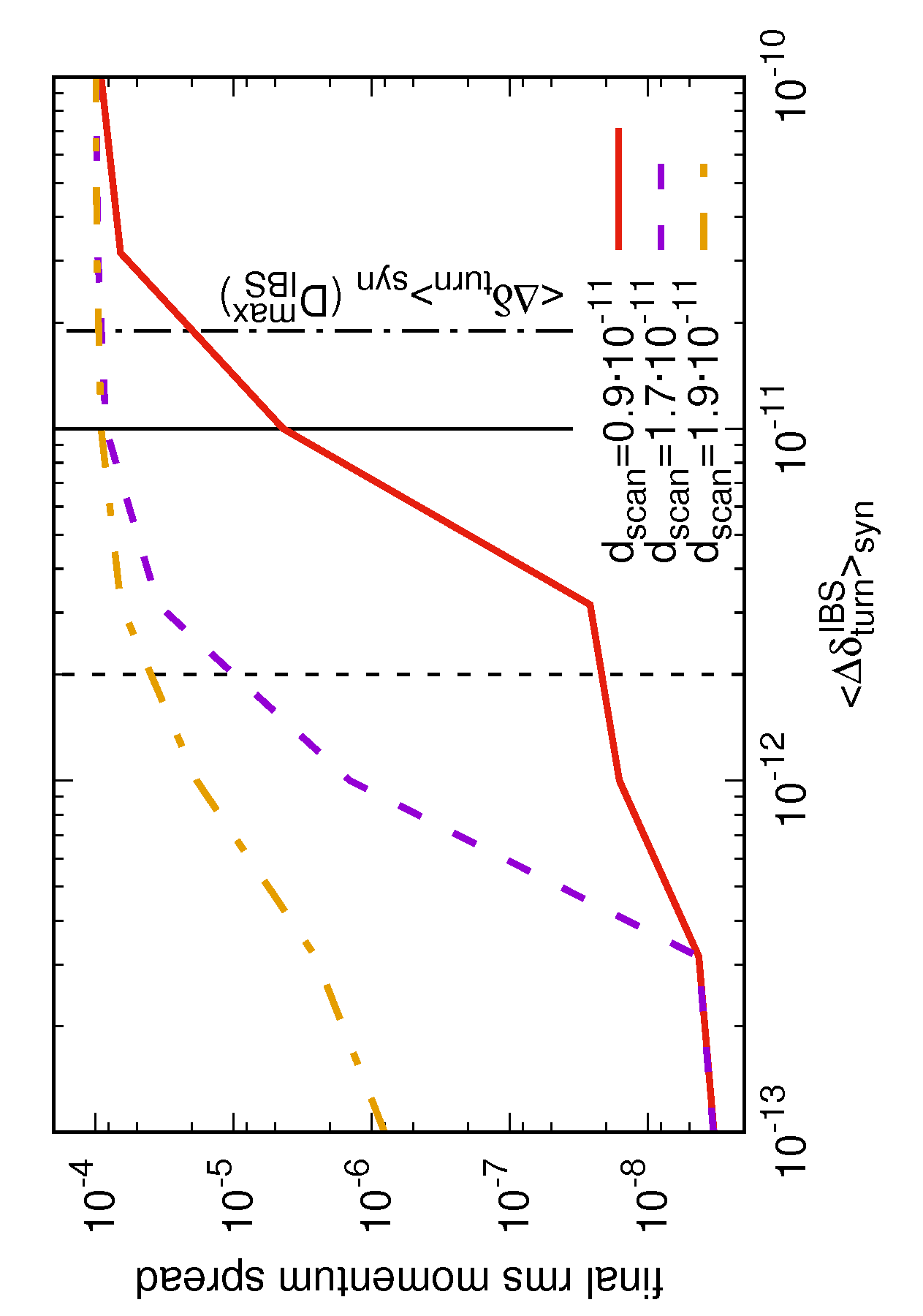}
	\caption{Simulation results of  the final rms momentum spread for different strength of IBS and different scan speeds of the laser resonance for the cooling of $\mbox{Ti}^{19+}$ ions in SIS100. The maximum IBS strength $D_{IBS}^{max}$ is marked, which corresponds to the theoretical limit of $d_{scan}=0$. Besides the maximum IBS strength is marked for $d_{scan} = 1.7\cdot 10^{-11}$ and $d_{scan}= 0.9\cdot 10^{-11}$. The maximum IBS strength for $d_{scan} = 1.9\cdot 10^{-11}$ is at $\left< \Delta \delta_{turn}^{IBS}\right>_{syn} = 0$.}
	\label{fig:cwIBS}
\end{figure}

During the laser cooling process, when the rms momentum spread of the bunch is much larger than the width of the laser force ($\delta_{rms} \gg \Delta_{fwhm}$), the cw laser force produces non-Gaussian profiles of the momentum distribution. For these bunch shapes the conventional Gaussian heating rate of diffusion is not valid anymore. The crucial process for the success of the cooling process is the broadening of the dense ring at the edge of the bunch in phase space (see fig. \ref{fig:scanSuccess}). The condition for a successful cooling process (see eq. \ref{equ:dscan}) has to be extended by an IBS heating term,

\begin{align}
	d_{scan}^{max} \approx \left<\Delta \delta^{cw}_{turn}\right>_{syn} - \left<\Delta \delta_{turn}^{IBS}\right>_{syn} \label{equ:dscanIBS}
\end{align}
where $\left<\Delta \delta_{turn}^{IBS}\right>_{syn}$ is the broadening of the dense ring in phase space in units of relative momentum per turn averaged over the synchrotron motion. The maximum scan speed $d_{scan}^{max}$ is now equal to the difference of the cooling by the laser and the heating by IBS. The broadening of the dense ring is calculated by the conventional IBS growth rate $\tau^{-1}_{IBS}$ assuming a Gaussian cross section of the dense ring,

\begin{align}
	\left<\Delta \delta^{IBS}_{turn}\right>_{syn} &= \delta_{ring}\left(e^{T_{rev}\cdot\tau_{IBS}^{-1}}-1\right)\cdot \frac{2}{\pi}\\
	\tau^{-1}_{IBS} &= \frac{D_{IBS\, 0}}{\delta_{ring}^2}
\end{align}
where $\delta_{ring}$ is the rms momentum spread of the dense ring at $z=0$, $D_{IBS\, 0}$ the initial diffusion coefficient and the factor $\frac{2}{\pi}$ arises by averaging the diffusive heating over the synchrotron motion. The fwhm of the dense ring is assumed to be equal to the laser force which result in:

\begin{align}
	\delta_{ring}\approx\frac{\Delta_{fwhm}}{2\sqrt{2\log{2}}}
\end{align}
Using eq. \ref{equ:dscanIBS} the maximum diffusion coefficient of the initial ion bunch respectively the highest ion intensity can be calculated for the limiting case $d_{scan}=0$ by

\begin{align}
	D_{IBS}^{max} \approx \frac{\left<\Delta \delta^{cw}_{turn}\right>_{syn}\cdot \Delta_{fwhm}\cdot \pi}{4\sqrt{2\log{2}}\cdot T_{rev}}. \label{equ:IBSmax}
\end{align}

Figure \ref{fig:cwIBS} shows the success of the cooling process for different strengths of IBS and different scan speeds of the laser resonance for the cooling process of $\mbox{Ti}^{19+}$ ions in SIS100. The maximum IBS strengths for the different laser scan speeds are marked by vertical lines. Higher ion intensities can be cooled with a slower scan of the laser resonance, but reducing the scan speed significantly below the value, that is calculated without intensity effects (see eq. \ref{equ:dscan}), the cooling process of the bunch with marginal higher intensity requires much more time. Note that the analytic estimations assume a fast synchrotron motion of the ions and might be inaccurate for low synchrotron frequencies (see section \ref{S:3.1}).

\subsection{Space charge effects for cw Laser} \label{S:4.2}

\begin{figure}[ht]
	\centering
	\includegraphics[width=.47\textwidth]{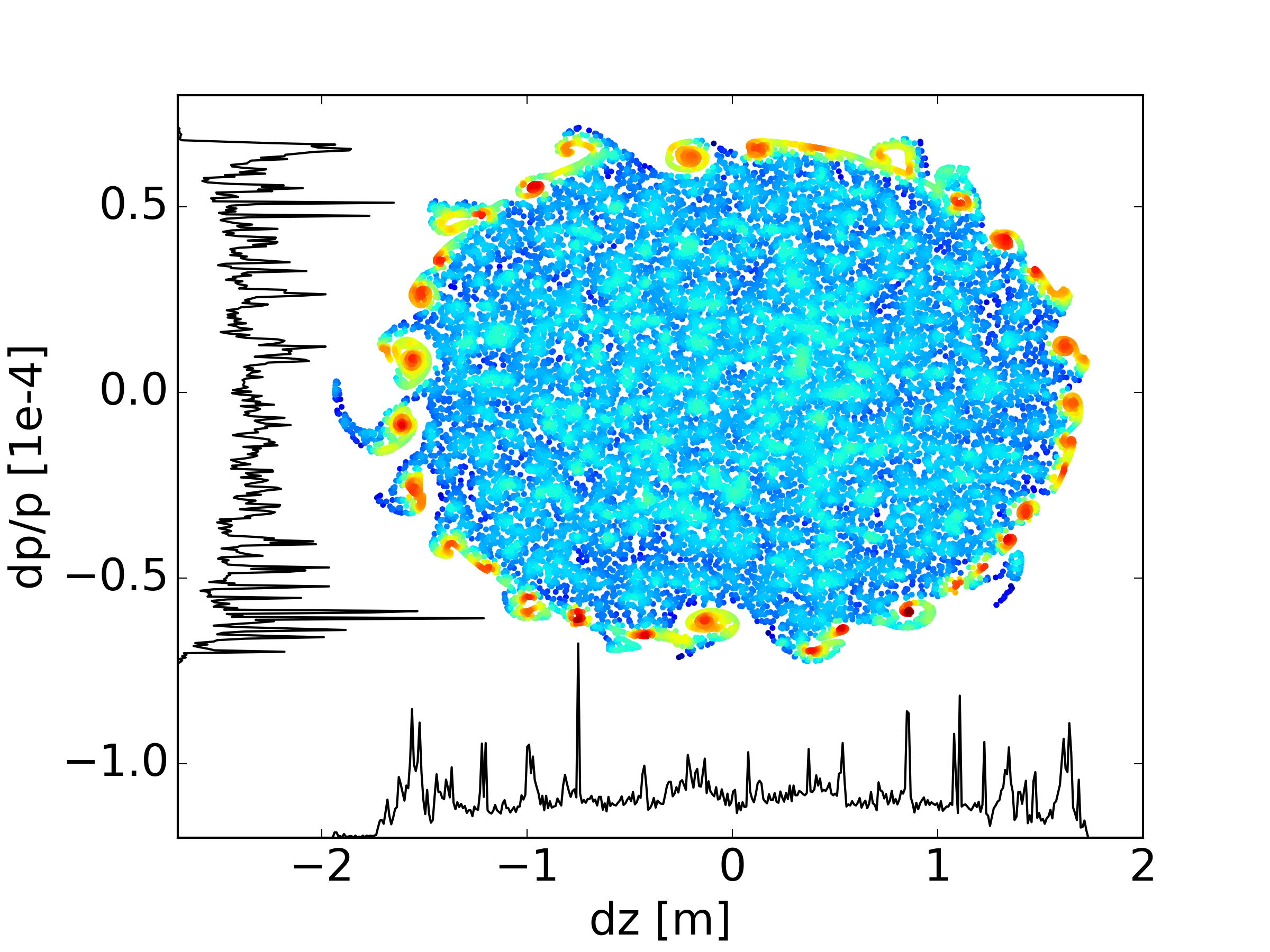}
	\caption{Phase space distribution of a bunch during the cooling process with a cw laser. SC leads to an instability of the dense ring, that is created by the narrow band cw laser force.}
	\label{fig:mwPhaseSpace}
\end{figure}

\begin{figure}[ht]
	\centering
	\includegraphics[width=.6\textwidth]{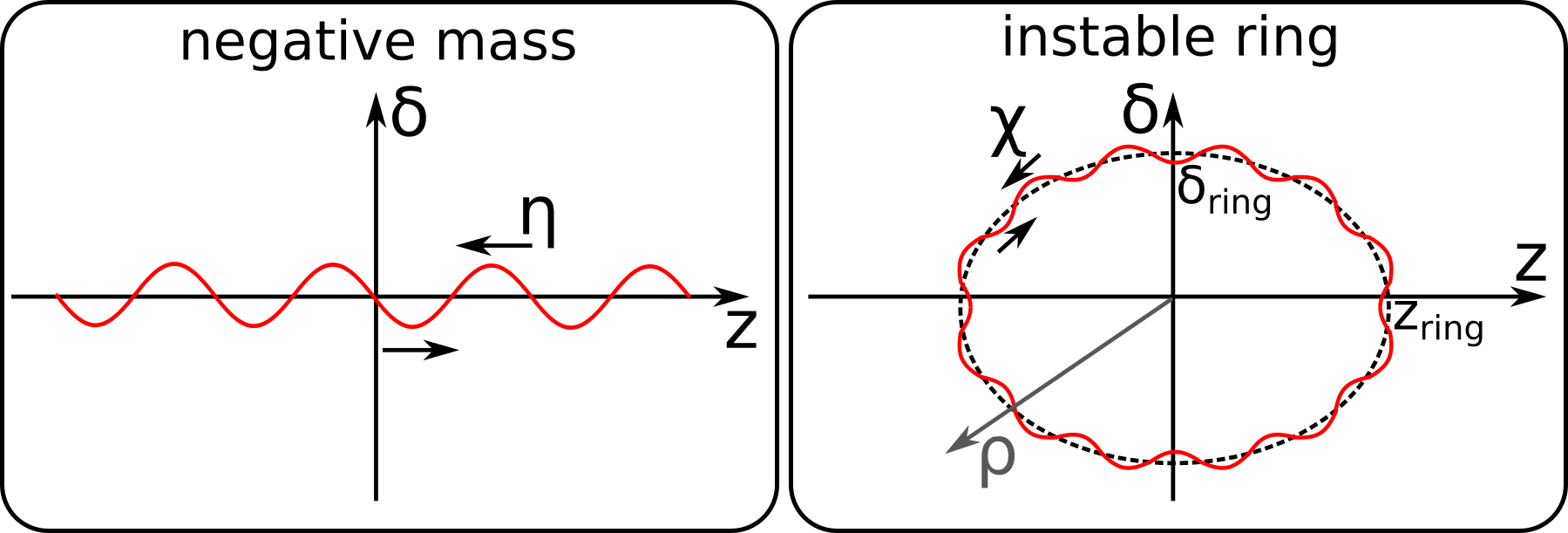}
	\caption{The Sketch illustrates the similarities of the negative mass instability and the instability of the ring. The SC induced synchrotron tune shift $\chi$ acts like the slip factor $\eta$ in the negative mass instability and leads to micro bunches.}
	\label{fig:mwsketch}
\end{figure}

\begin{figure}[ht]
	\centering
	\includegraphics[angle=-90, width=.47\textwidth]{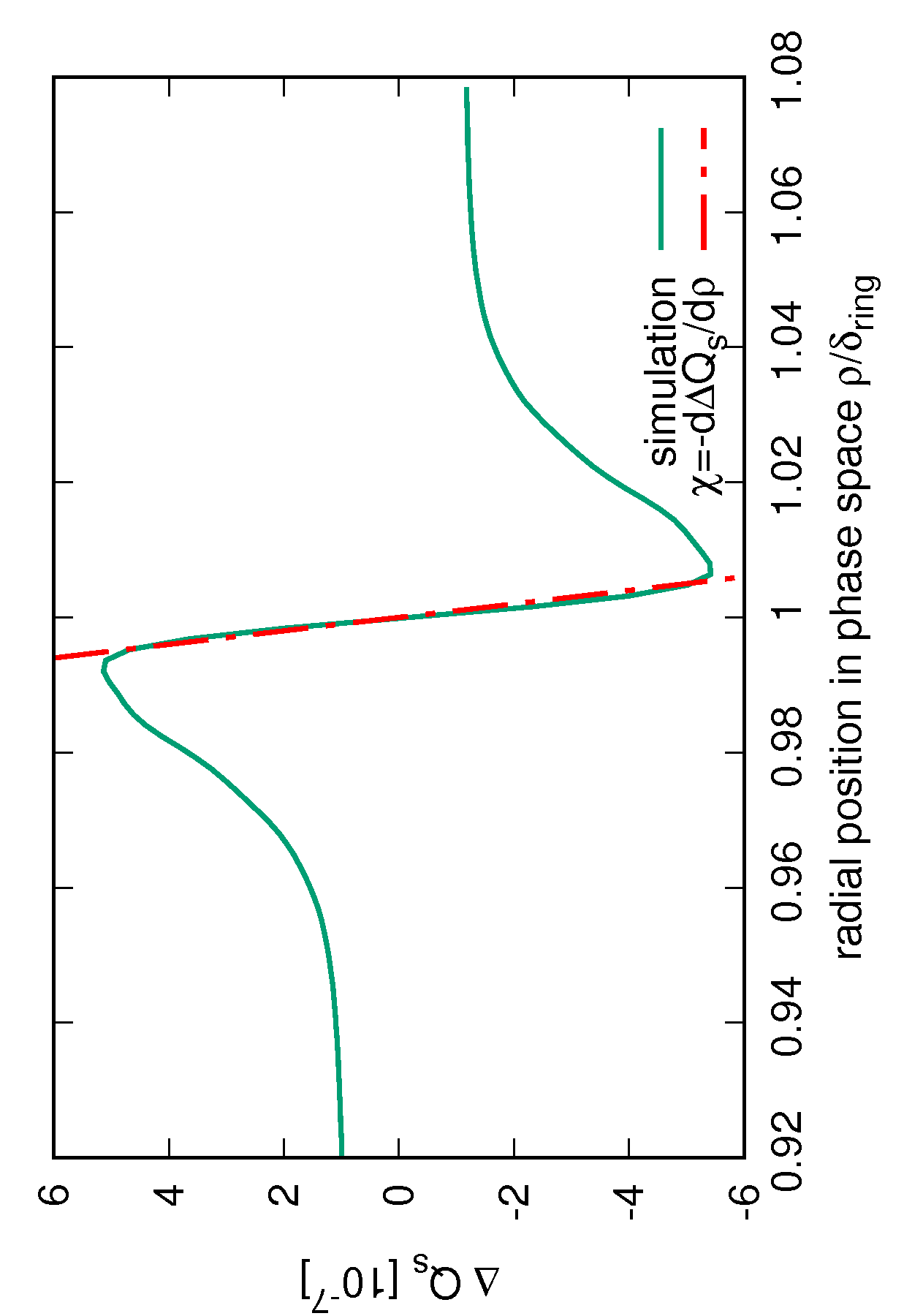}
	\caption{SC induced synchrotron tune shift that leads to an instability of the dense ring in phase space. The tune shift is given for an initial synchrotron tune of $Q_s = 3.5\cdot 10^{-4}$.}
	\label{fig:mwTuneShift}
\end{figure}

Beside the limitation of the shortest bunch length, SC leads to a collective instability during the cooling process with a cw laser. The dense ring in phase space, that is created by the narrow band cw laser force (see fig. \ref{fig:scanSuccess}), becomes instable and creates micro bunches, like shown in fig. \ref{fig:mwPhaseSpace}. The origin of this instability is similar to the negative mass instability in synchrotrons operating above transition energy (see ref. \cite{Lee2012}). Focusing on the fraction of the ring at $z=0$, ions with a higher absolute momentum deviation move less during one synchrotron period than ions with a lower absolute momentum deviation. In other words, ions of the dense ring with a higher oscillation amplitude have a lower synchrotron frequency compared to ions with a lower oscillation amplitude (see fig. \ref{fig:mwsketch}). For this instability the gradient of the synchrotron tune shift $\chi$ takes the role of the slip factor $\eta$ in the negative mass instability. Different synchrotron frequencies for different positions close to the ring cause an increase of an initially small perturbation of the ring, like illustrated in fig. \ref{fig:mwsketch}.  This tune shift is created by the SC fields of the peaks at the head and tail of the bunch (see fig. \ref{fig:scanSuccess}), and is exemplary shown in fig. \ref{fig:mwTuneShift}. The gradient of the synchrotron tune shift $\chi$ is given by:

\begin{align}
	\chi &= \left. -\frac{d \Delta Q_s}{d\rho}\right|_{\rho=\delta_{ring}}\\
	\rho(\delta, z) &= \sqrt{\delta^2+\left(\frac{\delta_{ring}}{z_{ring}}\cdot z\right)^2}
\end{align}
where $\rho$ is the radial position in phase space normalized to the positions of the dense ring in relative momentum $\delta_{ring}$ and the longitudinal position $z_{ring}$, neglecting the non linearity of the rf bucket (see fig. \ref{fig:mwsketch}).

\begin{figure}[ht]
	\centering
	\includegraphics[angle=-90, width=.47\textwidth]{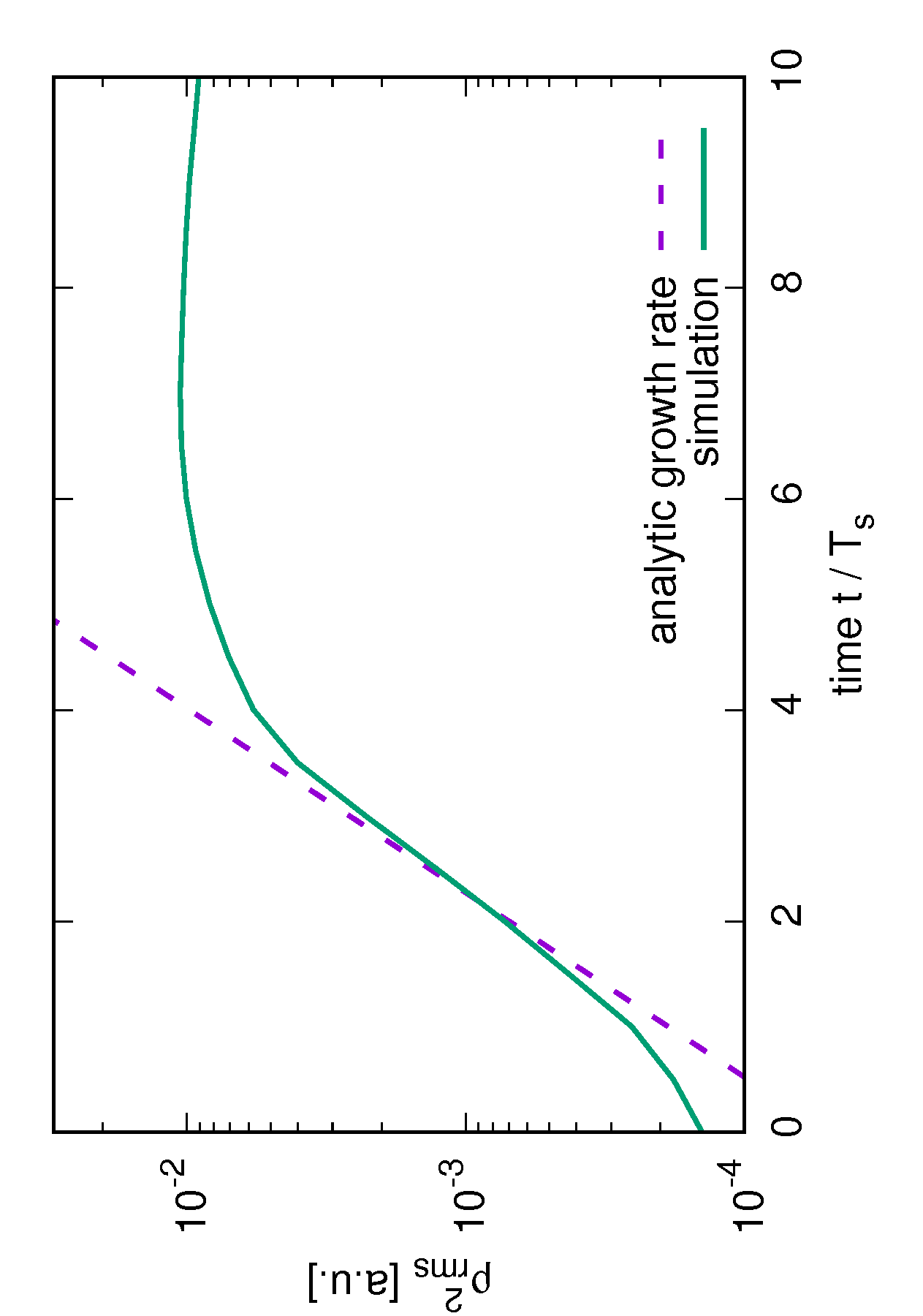}
	\caption{Rms width of the dense ring in phase space $\rho_{rms}$ during the growth of the instability over time normalized to the synchrotron period $T_s=T_{rev}/Q_s$. The wave vector of the initial perturbation is $k=60\;\frac{1}{m}$ and the synchrotron tune shift is shown in fig. \ref{fig:mwTuneShift}.}
	\label{fig:mwHamilton}
\end{figure}

Under the assumption of a short synchrotron period compared to the rise time of the instability, the Vlasov equation is solved for this ion distribution in \ref{A:3}. SC leads to an exponential growth rate of the perturbation

\begin{align}
	\omega = \pm ik\sqrt{\frac{q^2 g_{sc} N \chi}{8\pi\epsilon_0 m \gamma ^3 L_{acc}}}, \label{equ:mwrise}
\end{align}
where $k$ is the wave number of the perturbation on the dense ring in phase space in units of $1/m$. In fig. \ref{fig:mwHamilton} the growth rate is compared to simulation results of a ring in phase space with an initial sinusoidal perturbation like illustrated in fig. \ref{fig:mwsketch}. The growth rate of the rms width of the ring $\rho_{rms}$  agrees with eq. \ref{equ:mwrise} until Landau damping overcomes the instability. Despite the good agreement of the analytic description and the simulation, a precise threshold for a maximum ion intensity for laser cooling is difficult to define. The instability leads to a broadening and a density modulation of the ring, while the laser force only counteracts the broadening and does not affect the density modulation. However for typical SC strength the instability only leads to an additional heating beside IBS and does not in general prevent a successful cooling process.

Equation \ref{equ:mwrise} indicates an increase of the growth rate for perturbations with higher wave numbers. The SC field for a wave with the frequency above the SC cutoff is damped in the accelerator. Consequently the frequency of the wave with the shortest rise time is equal to the SC cutoff. For higher beam energies the SC cutoff increases ($f_{cutoff}\propto \beta \gamma$). Therefore this SC instability might be an issue for cooling of relativistic ion beams although this effect has not been noticed in laser cooling experiments so far.

\subsection{Pulsed Laser}

\begin{figure}[ht]
	\centering
	\includegraphics[angle=-90, width=.47\textwidth]{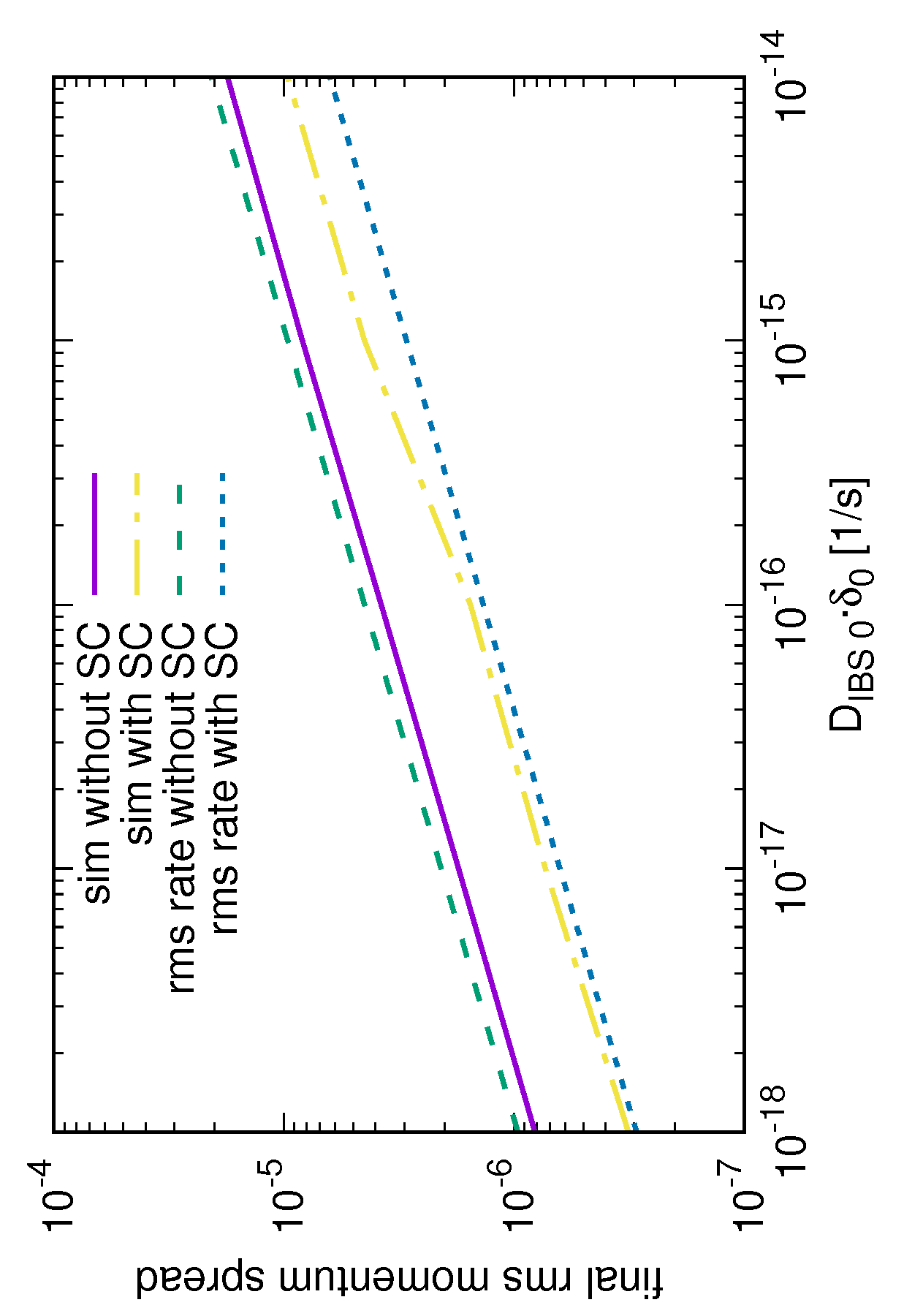}
	\caption{Simulation results of equilibrium state of IBS and cooling force of pulsed laser system for cooling of $\mbox{Ti}^{19+}$ ions in SIS100.}
	\label{fig:PulsedIbs}
\end{figure}
In contrast to the cooling process with a cw laser, the distribution function stays nearly Gaussian during the whole cooling process with a pulsed laser system. The cooling rate, calculated by eq. \ref{equ:pulseFcool}, can be compared to the IBS heating rate that is evaluated by

\begin{align}
	\tau_{IBS}^{-1} =\frac{D_{IBS}}{\delta^2_{rms}}.
\end{align}
The diffusion coefficient reaches its maximum at the shortest possible bunch length (see eq. \ref{equ:zMin})

\begin{align}
	\hat D_{IBS}&=\frac{D_{IBS\,0} \cdot \hat z_0}{\hat z_{equ}},
\end{align}
where $\hat z_0$ is one half of the initial bunch length. Including the effect of SC, the equilibrium state results in

\begin{align}
	\delta_{rms} = \sqrt{\frac{2D_{IBS\,0} T_{rev} \sigma^2_\delta}{\left<\Delta\delta^{LF}\right> e^{-\frac{1}{2}\left(\frac{\delta_{LPos}}{\sigma_\delta}\right)^2}\delta_{LPos}}\frac{\hat z_0}{\hat z_{equ}}} \propto N_p^{1/3}.
\end{align}
The rms momentum spread is proportional to the number of ions per bunch to the power of $1/3$. Due to the limited reduction of the bunch length, SC reduces the equilibrium rms momentum spread of IBS and the laser force. The analytic calculations agree with the simulations as shown in fig. \ref{fig:PulsedIbs}. The small deviations arise due to the noise of the SC solver, that adds an additional diffusive heating to the ions.

\section{Prospects for Laser Cooling of Relativistic Beams}
\label{S:5}

\begin{table}[ht]
\centering
\begin{tabular}{llllllll}
& Ion & $T_{rev}$ & $L_{interact}$ & $\delta_{rms\,0}$ & $d_{beam}$ & $\gamma$ & $L_{bunch}$\\
\hline
ESR & $^{12}\mbox{C}^{3+}$ & $0.8\,\mu s$ & $25\, m$ & $10^{-5}$ & $3\, mm$ & $1.13$ & $4\, m$\\
\hline
SIS100 & $ ^{48}\mbox{Ti}^{19+}$ & $3.6\,\mu s$ & $26\,m$ & $10^{-4}$ & $10\, mm$ & $8.50$ & $100\, m$
\end{tabular}
\caption{Comparison of accelerator parameters of ESR and SIS100 for the two examples of the laser cooling process.}
\label{tab:ESRSIS}
\end{table}

\begin{table}[ht]
\centering
\begin{tabular}{lllllll}
& $\lambda_L$ & $P_{L}$ & $n_{scat}$ & $\Delta \delta^{LF}$ & $\Delta_{fwhm}$ & $\rho_{excit}^{pulsed}$\\
\hline
$^{12}\mbox{C}^{3+}$ & $256\,nm$ & $20\, mW$ & $7.5$ & $1.5\cdot 10^{-9}$ & $5.8\cdot 10^{-8}$ & $0.278$\\
\hline
$ ^{48}\mbox{Ti}^{19+}$ & $512\, nm$ & $5\, W$ & $4.7$ & $9.3\cdot 10^{-10}$ & $3.9\cdot 10^{-8}$ & $8\cdot 10^{-5} $
\end{tabular}
\caption{Comparison of the laser parameters for the two examples of the laser cooling process.}
\label{tab:ESRSISLaser}
\end{table}

In order to evaluate the ability of laser cooling at relativistic beam energies, an example of a cooling process in SIS100 is compared to the reference scenario of $\mbox{C}^{3+}$ ions in ESR, that was already studied experimentally (see e.g. ref. \cite{bussmann2007}). The important parameters of the accelerators are shown in tab. \ref{tab:ESRSIS} and the expected laser forces in tab. \ref{tab:ESRSISLaser}. Due to the possibility of using an electron cooler in ESR the initial momentum spread is already one order of magnitude below the expected momentum spread in SIS100. The electron cooler also reduces the transverse diameter of the ion beam $d_{beam}$ that defines the diameter of the laser beam. For larger transverse dimensions of the particle beam, the saturation of the atomic transition requires more laser power. In order to provide sufficient laser power, the cooling process of $\mbox{Ti}^{19+}$ ions is studied with a laser wavelength of $512\,nm$. The pulsed laser system is assumed to be synchronized to the revolution frequency of the ions ($f_{rep} = f_{rev}$) and to  have the same average intensity as the cw laser system. The pulse length is adjusted the way, that the standard deviation of the laser force is 0.6 of the rms momentum spread of the initial particle distribution ($\sigma_\delta = 0.6\cdot \delta_0$) and is positioned at $\delta_{LPos} = 0.8\cdot \delta_0$ that were found to be good values for an efficient cooling process. For the cooling of $\mbox{C}^{3+}$ ions the pulsed laser intensity is not high enough to achieve the ideal laser force ($\rho_{excit}^{pulsed}=1$) but the intensity is sufficient to be as fast as the cooling scheme of the cw laser. In case of $\mbox{Ti}^{19+}$ ions the laser intensity is too low in order to cool all ions simultaneously with a pulsed laser. For this example the pulsed laser system could only be used to support the cw laser.

\begin{figure}[ht]
\centering
\includegraphics[width=.47\textwidth]{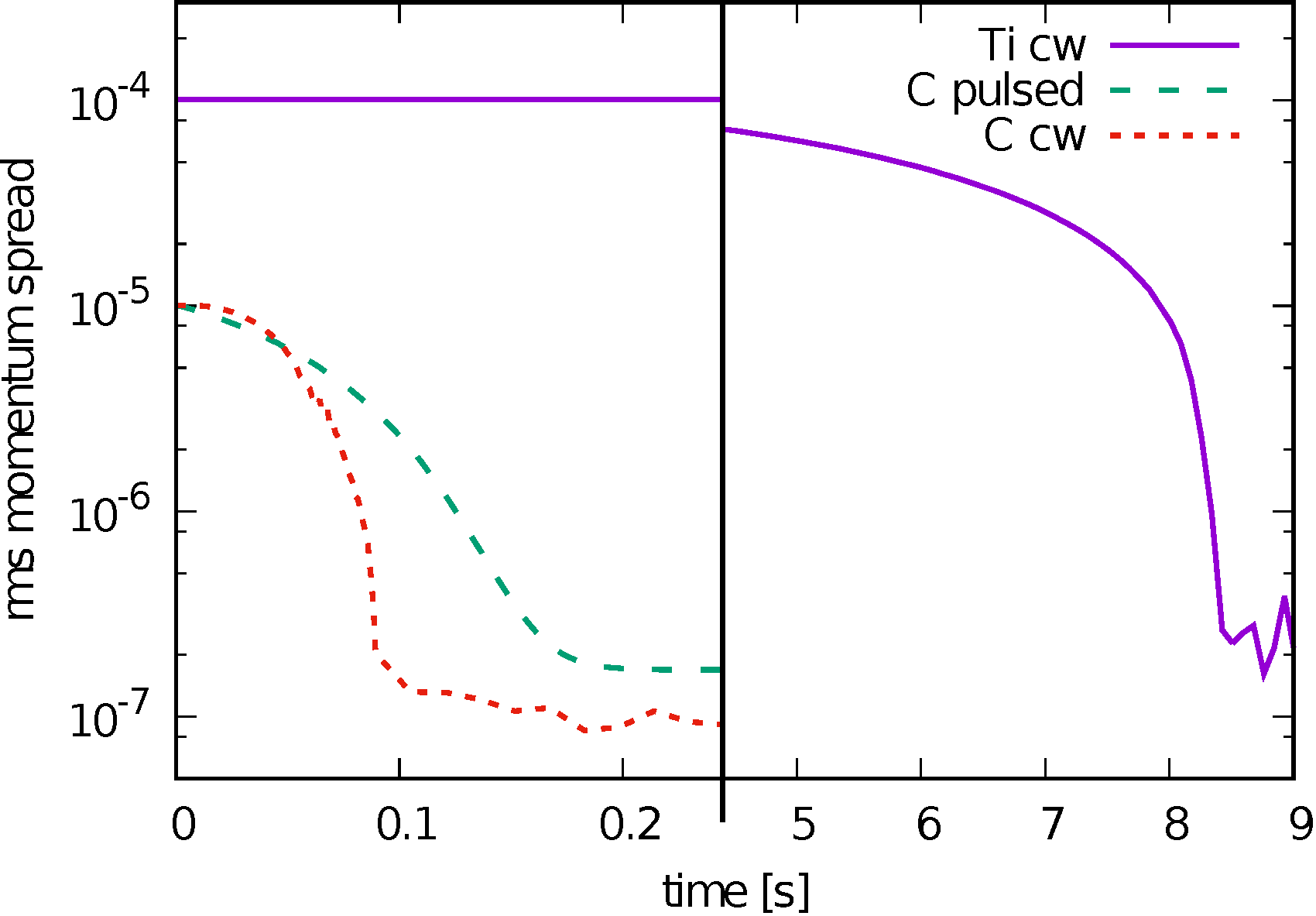}
\caption{Rms momentum spread over time for the cooling process of $\mbox{C}^{3+}$ and $\mbox{Ti}^{19+}$ ions. The cooling scheme of a cw and a pulsed laser system are compared for $\mbox{C}^{3+}$ ions. The required cooling time in SIS100 is much longer compared to the cooling time in ESR.}
\label{fig:ESRSIS}
\end{figure}

The simulation results of the rms momentum spread over time are shown in fig. \ref{fig:ESRSIS}. The applied scan speeds of the cw laser resonance for the simulations and the analytic predictions (see eq. \ref{equ:dscan}) are shown in tab. \ref{tab:ESRSISres}. For $\mbox{Ti}^{19+}$ ions the scan speed in simulation is faster than the analytic prediction because the slow synchrotron motion of the ions in the bucket violates the assumptions of the analytic formula (for more information see section \ref{S:3.1}) and therefore the formula does not give an accurate result. For the cooling process of $\mbox{C}^{3+}$ ions the scan speed in simulation and the analytically calculated maximum scan speed are equal. The necessary reduction of the scan speed due to the presence of intensity effects (see eq. \ref{equ:dscanIBS}) is compensated by a small effect of the slow synchrotron motion described in section \ref{S:3.1}.

\begin{table}[ht]
\centering
\begin{tabular}{lll}
& $^{12}\mbox{C}^{3+}$ & $^{48}\mbox{Ti}^{19+}$\\
\hline
$d_{scan}^{max}$ analytic & $2\cdot 10^{-10}$ & $2\cdot 10^{-11}$\\
\hline
$d_{scan}$ sim & $2\cdot 10^{-10}$ & $1\cdot 10^{-10}$\\
\hline
$N_p^{max}$ analytic & $6.7\cdot 10^6$ & $6.8\cdot 10^7$\\
\hline
$N_p$ sim & $1.5\cdot 10^6$ & $10^7 $\\
\hline
$L_{equ}$ sim & 1.4 m & 3.6 m 
\end{tabular}
\caption{Simulation results compared to analytic predictions of the scan speed and the maximum ion intensity per bunch for a harmonic number $h=20$ in ESR and $h=8$ in SIS100. }
\label{tab:ESRSISres}
\end{table}

Despite similar values for the strength of the laser force and the laser scan speed for both examples, the required cooling time in SIS100 is much higher compared to the cooling time in ESR (see fig. \ref{fig:ESRSIS}). The main reason for the difference in cooling time is given by the initial momentum spread, that scales linear with the time for the cooling process with a cw laser system. In addition the revolution time is about five times higher in SIS100 compared to ESR whereas the length of the cooling section has similar length. The initial momentum spread and the difference in revolution frequency cause a factor of $45$ in cooling time between ESR and SIS100 (see eq. \ref{equ:Tcoolcw}). The simulated cooling time of $\mbox{C}^{3+}$ ions in ESR is much shorter compared to the experimental results, because simulations assume perfect conditions of the interaction section and the scan speed of the laser resonance is adjusted to the highest possible value, that was not the case in experiments.

The intensity limits of the simulations are compared to the analytically calculated limits in tab. \ref{tab:ESRSISres}. The analytic prediction can only be calculated for $d_{scan} \rightarrow 0$ (see eq. \ref{equ:IBSmax}) because the slow synchrotron motion violates the assumptions for an accurate calculation of the scan speeds. Therefore the analytic predictions overestimate the intensity limits in the simulations for a scan speed $d_{scan}>0$. The simulated intensity limit for this example is only valid for this particular scan speed of the laser resonance and can be increased by changing the scan speed like described in section \ref{S:4.1}. Besides the heating of IBS, the presence of SC causes an instability, as described in section \ref{S:4.2}. But in both cases the cooling process with a cw laser was disturbed by IBS for higher ion intensities.

\section{Conclusion}
\label{S:6}
The laser force for relativistic ions in storage rings or synchrotrons was calculated in detail for cw and pulsed laser systems using the optical Bloch equations. For Li-like ions and saturated transitions the obtained relative cooling force is very similar. However, at higher beam energies the required laser intensity increases strongly, which makes it difficult to provide sufficient laser power for saturation. The cooling processes for a cw laser system and a pulsed laser system are compared. 

Assuming that sufficient laser power is available, both schemes are competitive in terms of cooling time and beam intensity limitations. 

However, the cooling procedure 
and the particle dynamics of the cooling processes with a cw or a pulsed laser have major differences. Cooling with a cw laser system leads to ultra low momentum spreads but creates pronounced peaks in the longitudinal beam profile. A reduction of the synchrotron tune, more precisely the reduction of the rf force with respect to the laser force, reduces the required cooling time. The reduction is beneficial until the laser force is as strong as the rf force. A major difficulty is the correct choice of the scan speed of the laser resonance in phase space, which depends on the strength of the laser force, the ion intensity and the depth of the rf potential. In contrast, the cooling process with a pulsed laser does not require a scan of the laser frequency, which makes the cooling process much easier to handle. In addition, the beam profile stays nearly Gaussian during the whole cooling process and the particle dynamics can be well described by rms rate equations.

We show that the minimum bunch length is limited by space charge while the lowest momentum spread is defined by the equilibrium of the laser force and the heating due to intrabeam scattering. The simulation of the cooling process with a cw laser shows a space charge induced instability, that might be crucial for cooling of highly relativistic ion beams, because the cutoff frequency of the space charge impedance increases with energy.

The comparison of two cooling scenarios, $\mbox{C}^{3+}$ in the ESR and $\mbox{Ti}^{19+}$ in the SIS100, shows that for similar cooling forces the required cooling time is much longer in the SIS100 case. The higher initial momentum spread and the lower revolution frequency are the main reasons for this. The good agreement of simulations and analytic predictions motivates us to transfer the results to other possible atomic transitions.

\section{Acknowledgments}

The authors would like to thank Michael Bussmann (HZDR) for valuable discussions. This project is supported by the BMBF, project number 05P15RDFA1.

\appendix

\section{Simulation Model} \label{A:1}

For simulation studies of the cooling process a longitudinal tracking code with the particle-in-cell (PIC) method is applied. The simulation code includes the diffusive heating of intrabeam scattering (IBS) by the local diffusion model for non-Gaussian beam profiles as described in ref.  \cite{betacool2008} and space charge (SC) implemented like described in ref. \cite{BoineFrankenheim2005}. For the laser ion interaction two different models for the cw laser and one for the pulsed laser are implemented:
\subsection{Statistical cw Laser Force}

\begin{figure}[ht]
	\centering
	\includegraphics[angle=-90, width=.4\textwidth]{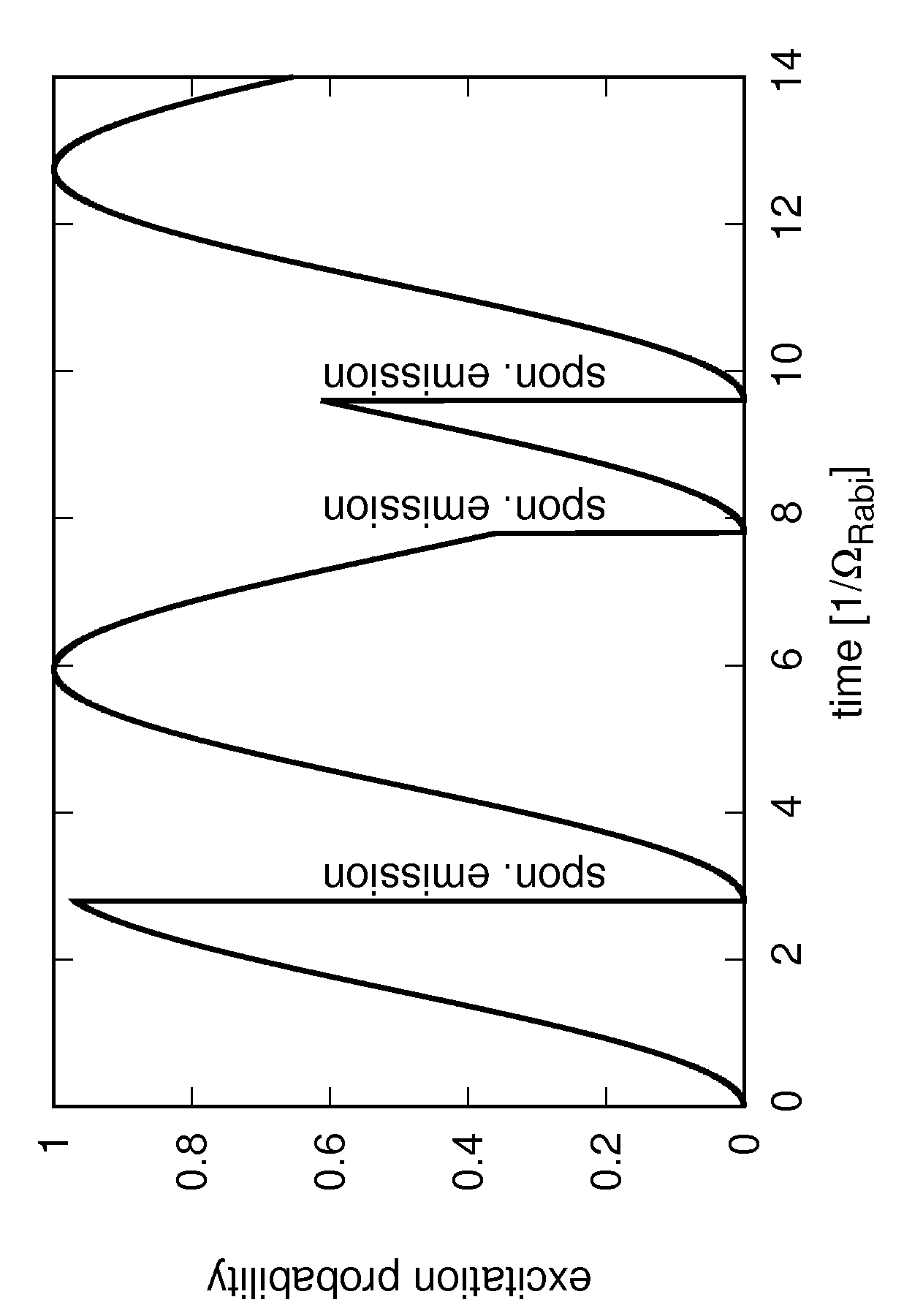}
	\caption{Example of excitation probability of a single ion. The excitation probability oscillates and is interrupted by spontaneous emissions, whose probability of occurrence is equal to the excitation probability.}
	\label{fig:excitSingleIon}
\end{figure}
\begin{figure}[ht]
	\centering
	\includegraphics[angle=-90, width=.4\textwidth]{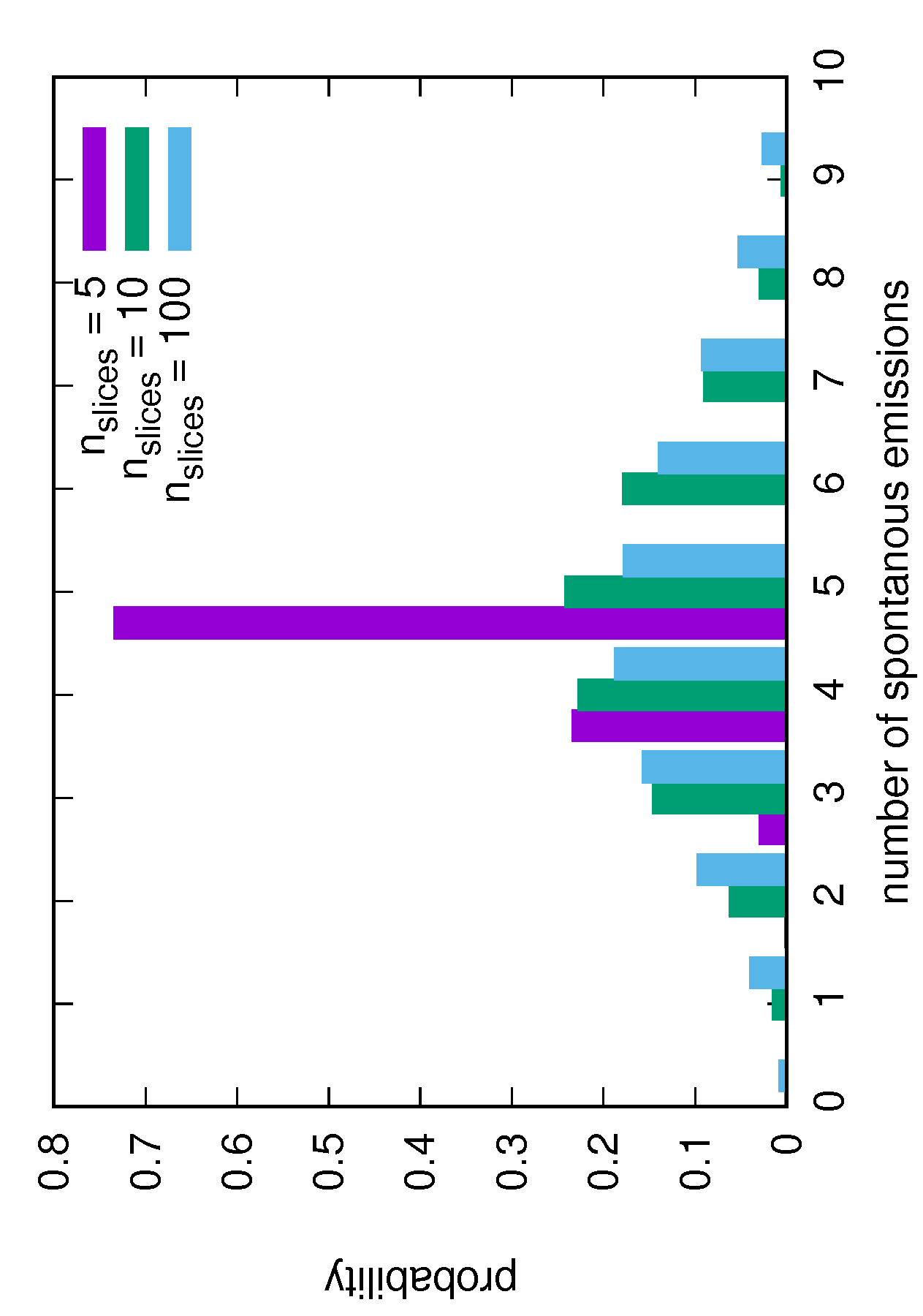}
	\caption{Histogram of the probability of different numbers of spontaneous emission for different amounts of slices. For $n_{slices}=5\approx n_{scat}^{turn}$ the statistical distribution is not well described whereas for twice the amount the distribution has already converged sufficiently. }
	\label{fig:sponEmis}
\end{figure}

The laser ion interaction consists of three effects given by the stimulated emission, the spontaneous emission and the absorption of a photon. Neglecting the spontaneous emission the excitation probability oscillates for each ion with the Rabi frequency ($\Omega_{rabi} = \sqrt{I(t)/2 I_S} \cdot 1/\tau_{se}$):

\begin{align}
	\rho_{ee}(t,\delta)=\frac{\Omega_{Rabi}^2}{\Omega_{Rabi}'^2(\delta)}\cdot \sin ^2\left(\frac{\Omega_{Rabi}'(\delta)\cdot t}{2}\right) \label{equ:singleIonRabi}\\
	\Omega_{Rabi}'(\delta)=\sqrt{\Omega_{Rabi}^2+\zeta^2\cdot(\delta-\delta_{LPos})^2}.
\end{align}
The symbols are explained in detail in sec. \ref{S:2}. The probability of a spontaneous emission is equal to the excitation probability. If a spontaneous emission take place, the ion receives the recoil of the emitted photon, the excitation probability drops immediately down to $\rho_{ee}=0$ and continues with oscillations (see example in fig. \ref{fig:excitSingleIon}). The mathematical description of the Rabi oscillation including the spontaneous emission would need to include the basis functions of all possible spontaneous emissions but the problem can be solved by using the Monte Carlo method (see ref. \cite{molmer1993}). However the simulation of laser cooling does not require the exact description of the laser ion interaction because the momentum change of the ions due to spontaneous emissions is the only relevant quantity for the cooling process. Therefore the model has to reproduce correctly the number of spontaneous emissions in a certain time interval and the momentum change for each spontaneous emission. In order to reduce the computational effort and still use a realistic description of the interaction, the excitation probability of eq. \ref{equ:singleIonRabi} is averaged over time and the interaction region is divided into slices ($n_{slices}\geq n_{scat}^{turn}(\delta_{LPos})$). The probability of a spontaneous emission in each slice is given by:

\begin{align}
	\rho_{scat}(\delta) &= \int_0^{\frac{L_{interact}}{\gamma \beta c_0 \cdot n_{slices}}} \left<\rho_{ee}(t,\delta)\right>_t dt= \frac{n_{scat}^{turn}(\delta)}{n_{slices}} \label{equ:Fstat}
\end{align}
In each slice for each ion the probability is compared to a random number and if the random number is smaller, the momentum change of a spontaneous emission is applied to the ion (see eq. \ref{equ:potonkick}). The number of spontaneous emissions for different $n_{slices}$ is shown in a histogram in fig. \ref{fig:sponEmis} for the case of $\mbox{Ti}^{19+}$ ions in SIS100 (average amount of spontaneous emissions: $n_{scat}^{turn}=4.7$). The results indicate that choosing only $5$ slices the statistical occurrence of spontaneous emissions is not well described, whereas for $n_{slices}=10\approx 2\cdot n_{scat}^{turn}$ the distribution has already converged sufficiently.

In fig. \ref{fig:cwModel} the influence of the scan speed on the success of the cooling process, that is described in detail in sec. \ref{S:3.1} especially in fig. \ref{fig:scanLaser}, is shown for $n_{slices} = 5$ and $n_{slices} = 10$. Despite the distribution of the statistical occurrence of spontaneous emissions is not equal for both cases (see fig. \ref{fig:sponEmis}), the results of the cooling process are very similar. Due to the large number of spontaneous emissions acting on each ion during the cooling process, the detailed process of a single spontaneous emission is not important as long as the average momentum change over time and the strength of diffusive heating is correctly reproduced.

\subsection{Mean cw Laser Force}

\begin{figure}[ht]
\centering
\includegraphics[angle=-90, width=.4\textwidth]{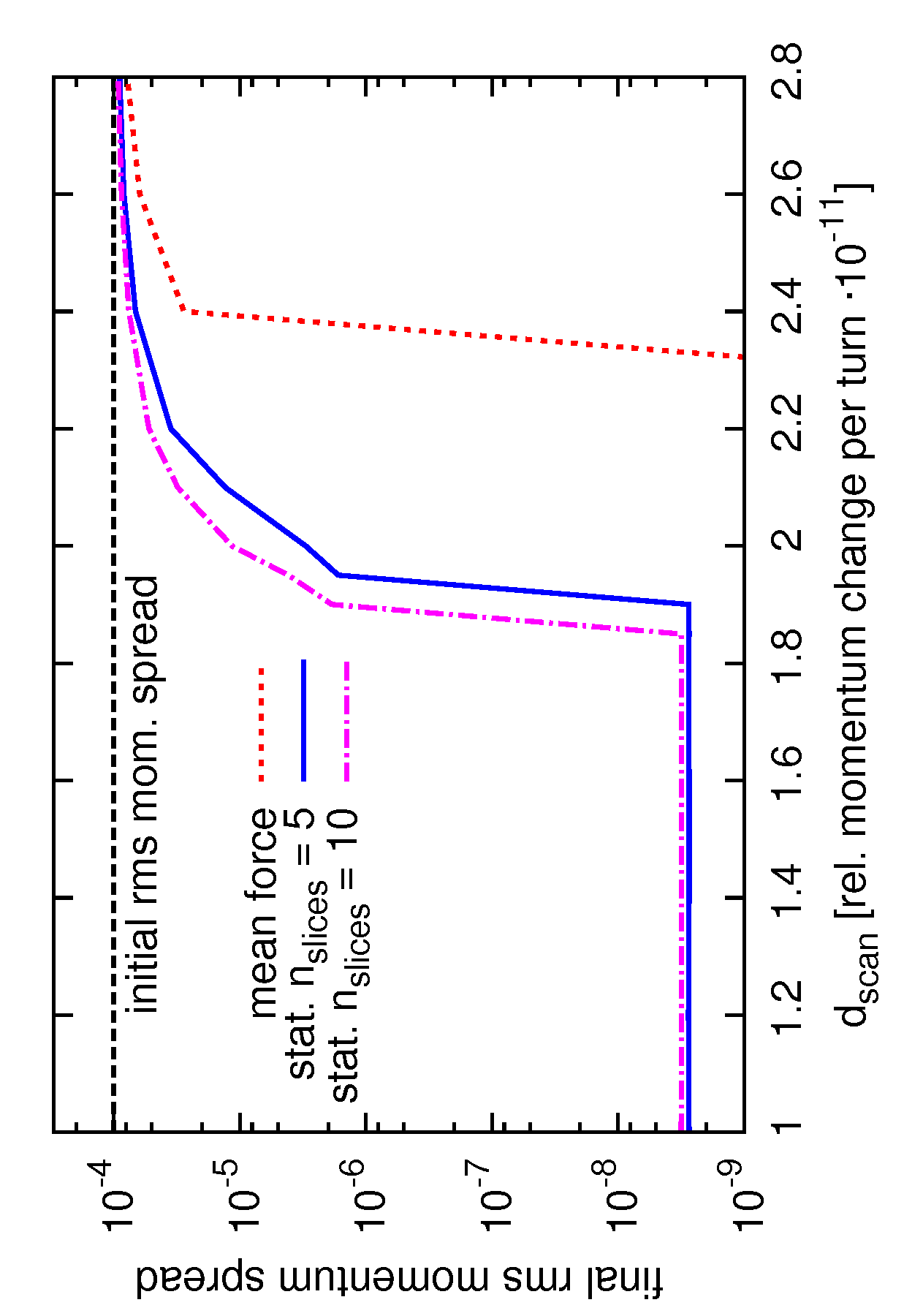}
\caption{Final rms momentum spread for different scan speeds of the laser force. The implementation of the mean laser force is compared to the statistical implementation with two different amounts of slices. The results of the two different simulations with the statistical implementation are very similar, whereas the implementation of the mean laser force shows some deviations.}
\label{fig:cwModel}
\end{figure}

For the implementation of the mean cw laser force, the statistical component of the laser ion interaction is neglected and the average momentum change per turn is applied to the ions. The momentum dependent average momentum change per turn is calculated by eq. \ref{equ:Favg} where the number of scattering events for the cw laser force is calculated by eq. \ref{equ:nscat}:

\begin{align}
	\Delta p_{turn}^{LF} = \left<\Delta p^{LF}\right>\cdot \frac{L_{interact}}{\gamma \beta c_0} \cdot \frac{1}{2\tau_{se}^{PF}}\frac{S}{1+S+(2 \zeta (\delta-\delta_{LPos}) \cdot \tau_{se}^{PF})^2}
\end{align}
Figure \ref{fig:cwModel} shows the differences of the implementation of the statistical and mean laser force. The implementation of the mean laser force does not include the laser induced  diffusion. The lack of diffusive heating enables to cool below the Doppler limit, which disagrees with the theory of Doppler laser cooling. On the other hand the maximum scan speed is in this example approximately $20\%$ higher due to the lack of the statistical process. However the evolution of the phase space distribution of a successful cooling process is very similar for both simulation models, especially for simulations including IBS and SC the details of the ion laser interaction become negligible. Therefore the simplified model is suitable for analytic descriptions and for most of the simulations.

\subsection{Pulsed Laser Force}
For the implementation of the pulsed laser force the probability of a scattering event in the interaction region, that is calculated by eq. \ref{equ:rhoScatPulsed} and is always lower or equal $1$, is compared to a random number. If the random number is smaller, the momentum change of a spontaneous emission is applied to the ion (see eq. \ref{equ:potonkick}). This model respects the random walk of the ions and covers the cooling as well as the diffusive component of the laser ion interaction.

\section{Excitation with a Pulsed Laser}\label{A:2}

\begin{figure}[ht]
	\centering
	\includegraphics[angle=-90, width=.47\textwidth]{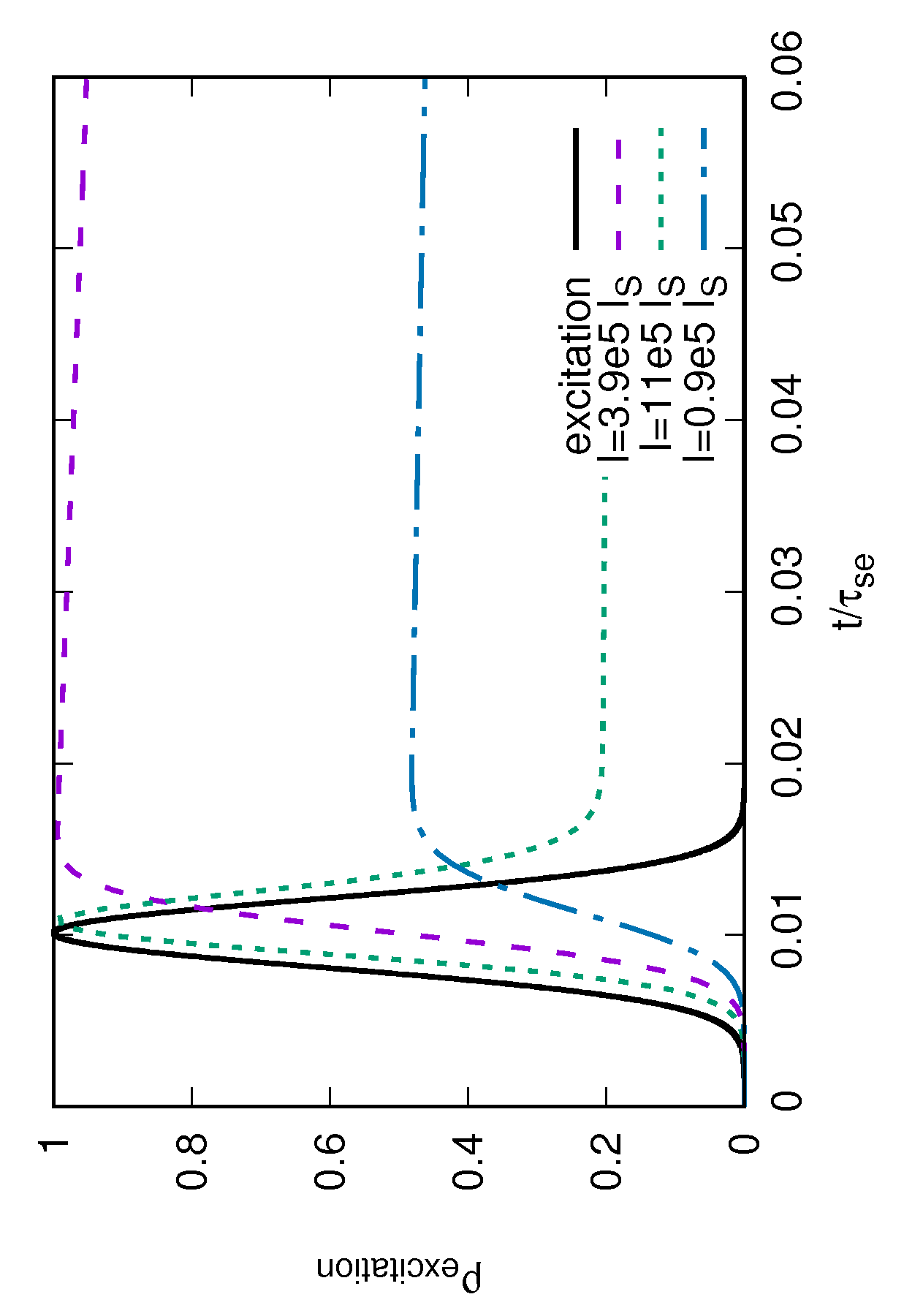}
	\caption{Excitation probability over time for different intensities of the pulsed laser system. During the laser pulse the ions perform Rabi oscillations. After the pulse the excitation slowly decays by spontaneous emission.}
	\label{fig:excitPulsed}
\end{figure}
The excitation of an ion ensemble with a pulsed laser with short laser pulses ($\sigma_t^{PF} \ll \tau_{se}$) can no long be described by the well known steady state solution. For the excitation with a short laser pulse, the optical Bloch equations (see ref. \cite{metcalf1999}) can be simplified for an infinite long lifetime of the excited state $1/\tau_{se}=0$ to

\begin{align}
	\frac{d^2(\rho_{ee}-\rho_{gg})}{dt^2}=-\Omega_{rabi}\cdot(\rho_{ee}-\rho_{gg}),
\end{align}
where $\rho_{gg}$ is the population probability of the ground state (all values are given in the PF).  With the constrains $\rho_{gg} + \rho_{ee} = 1$ and $\rho_{ee}=0$ at $t=0$ the excitation probability of the Rabi oscillator is

\begin{align}
	\rho_{ee}(t)=-\frac{1}{2}\cos\left(\Omega_{rabi} t\right) + \frac{1}{2}.
\end{align}
The intended state after the pulse is $\rho_{ee} = 1$. If the pulse energy is too low, some particles are still in the ground state whereas if the pulse energy is too high, the stimulated emission decreases the amount of ions in the excited state as shown in fig. \ref{fig:excitPulsed}. For analytic calculations the Gaussian pulse can be approximated by a rectangular pulse with the same peak intensity. The length of the corresponding rectangular pulse ($\tau_{rec\;pulse}=\sqrt{4\pi} \cdot \sigma_t$) is determined by the integral of the Rabi frequency over time that has to be equal for both pulse shapes.

\begin{align}
	\rho_{ee} =-\frac{1}{2}\cos\left(\Omega_{rabi} \cdot \tau_{rec\;pulse}\right) + \frac{1}{2} \overset{!}{=} 1 \label{equ:rabiosc}
\end{align}
The ideal laser peak intensity is

\begin{align}
	\frac{\hat{I}^{PF}}{I_{S}^{PF}} &= \frac{\hat{I}^{LF}}{I_{S}^{LF}} = \frac{\pi}{2} \cdot \left(\frac{\tau_{se}^{PF}}{\sigma_{t}^{PF}}\right)^2.
\end{align}
The analytically calculated peak intensity agrees with the results of the numerically solved Bloch equations.

\section{Space Charge Instability}\label{A:3}

\begin{figure}[ht]
	\centering
	\includegraphics[width=.5\textwidth]{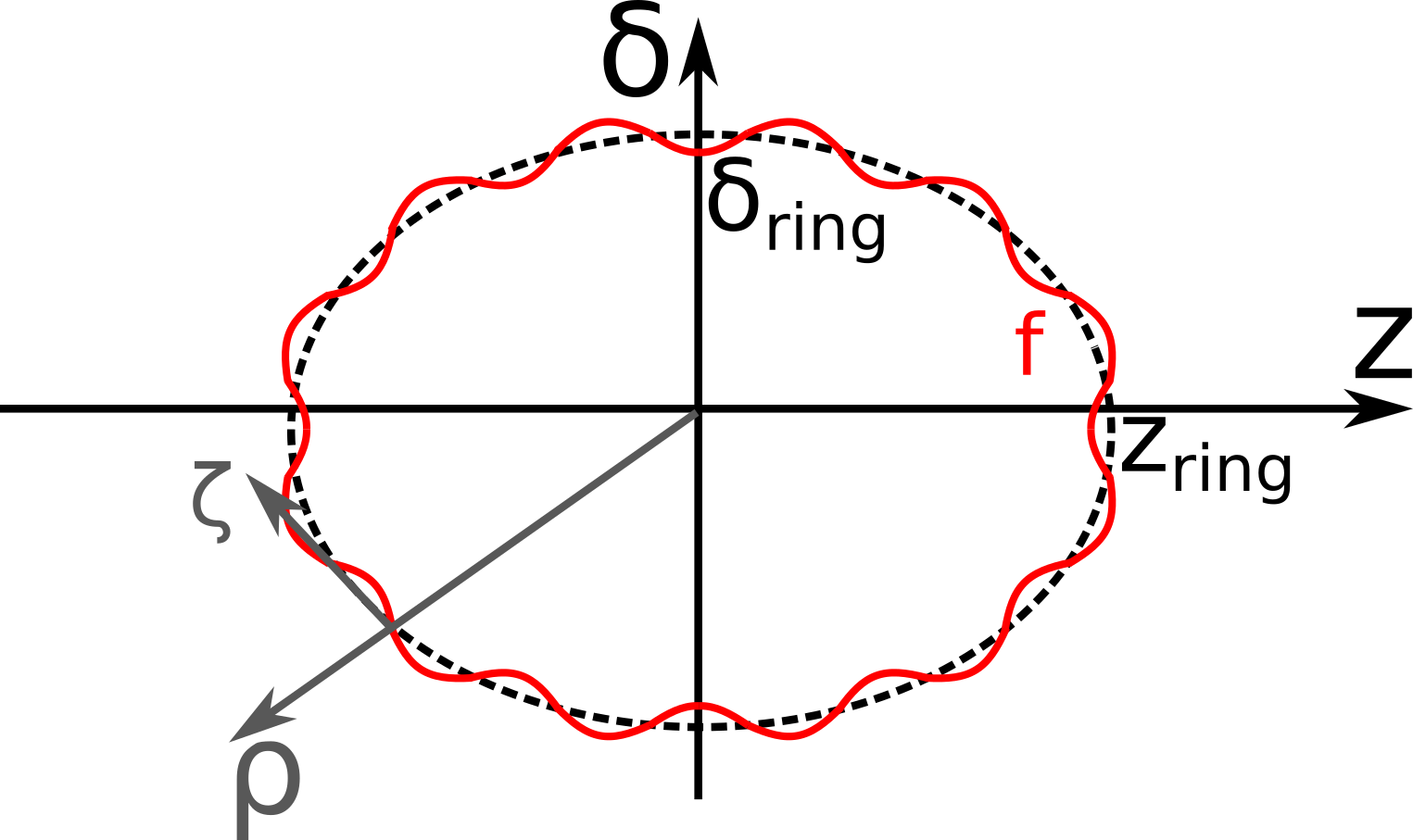}
	\caption{Sketch of the distribution function (red) and the polar like coordinate system (gray) in the conventional coordinate system (black). The polar like coordinate system ($\rho$, $\zeta$) rotates in phase space with the speed of the synchrotron motion.}
	\label{fig:mw_trans}
\end{figure}

The analytic description of the space charge instability of the dense ring in phase space is derived by solving Vlasov equation in the coordinate system, that is illustrated in fig. \ref{fig:mw_trans}.

\begin{align}
	\frac{d f}{d t}= \frac{\partial f}{\partial t} + \frac{\partial \rho}{\partial t} \cdot \frac{\partial f}{\partial \rho} + \frac{\partial \zeta}{\partial t}\cdot \frac{\partial f}{\partial \zeta} = 0 \label{equ:mw_vla}
\end{align}
$\rho$ and $\zeta$ describe the new coordinates, that are calculated by the conventional coordinates $\delta$ and $z$ by

\begin{align}
	\rho &= \sqrt{\delta^2 + \left(\frac{\delta_{ring}}{z_{ring}}\cdot z\right)^2}\\
	\zeta &= z_{ring} \cdot \left( \arctan \left(\frac{z_{ring}\cdot\delta}{z \cdot \delta_{ring}}\right) - 2\pi \frac{t}{T_{rev}}\cdot Q_s \right).
\end{align}
The coordinate $\rho$ describes the radial position in phase space in units of relative momentum. The coordinate $\zeta$ describes the angular position in phase space in units of meters. The origin rotates with the speed of the synchrotron motion in phase space. In order to solve eq. \ref{equ:mw_vla} in the polar like coordinate system, the derivations have to be averaged over the synchrotron motion.

\begin{align}
	\frac{\partial f}{\partial t} + \left< \frac{\partial \rho}{\partial t}\right>_{syn} \cdot \frac{\partial f}{\partial \rho} + \left<\frac{\partial \zeta}{\partial t} \right>_{syn} \cdot \frac{\partial f}{\partial \zeta} = 0 \label{equ:mw_vla2}
\end{align}
This simplification claims that the analyzed process is slow compared to the synchrotron motion. Following the derivation of the negative mass instability (see e.g. ref. \cite{Lee2012}) the distribution function $f$ including a small perturbation $f_1$ with a wave vector $k$ in units of $1/m$ is defined by:

\begin{align}
	f &= f_0 + f_1 \label{equ:mwdist}\\
	f_0 &= \frac{N_p}{2\pi z_{ring}} \cdot \delta\left(\rho - \delta_{ring}\right)\\
	f_1 &\propto e^{i(k \zeta - \omega t)}
\end{align}
The time derivative of the radial position $\rho$ is given by:

\begin{align}
	\frac{\partial \rho}{\partial t} &= \frac{\partial \delta}{\partial t}\cdot \frac{\partial \rho}{\partial \delta} = \frac{\partial \delta}{\partial t} \cdot \frac{\delta}{\delta_{ring}}\\
	&= \frac{\delta^2}{\delta_{ring}^2} \cdot \frac{q^2g_{sc}}{4\pi \epsilon_0 \gamma^2 p_0}\cdot i k n_1
\end{align}
After averaging over the synchrotron motion the derivative is independent of $\delta$ and $z$.

\begin{align}
	\left<\frac{\partial \rho}{\partial t}\right>_{syn} &= \frac{1}{2}  \cdot \frac{q^2g_{sc}}{4\pi \epsilon_0 \gamma^2 p_0}\cdot i k n_1 \label{equ:derRho}
\end{align}
The average of the time derivative of the angular position $\zeta$ is given by:

\begin{align}
	\left<\frac{\partial \zeta}{\partial t}\right>_{syn} &= \frac{\partial}{\partial t} \frac{2\pi t}{T_{rev}} \Delta Q_s\\
	&= -\frac{2\pi}{T_{rev}} \rho \cdot \chi \label{equ:derzeta}
\end{align}
Combining eq. \ref{equ:mw_vla2}, \ref{equ:derRho} and \ref{equ:derzeta} the growth of the perturbation in the distribution function (eq. \ref{equ:mwdist}) is calculated equivalent to the negative mass instability (see e.g. ref. \cite{Lee2012}). The growth rate is given by eq. \ref{equ:mwrise} and agrees with simulation results.





\bibliographystyle{ieeetr}
\bibliography{laserCooling.bib}

\begin{thebibliography}{10}

\bibitem{Phillips1998}
W.~D. Phillips, ``Nobel lecture: Laser cooling and trapping of neutral atoms,''
  {\em Rev. Mod. Phys.}, vol.~70, pp.~721--741, Jul 1998.

\bibitem{schroeder1990}
S.~Schr\"oder, R.~Klein, N.~Boos, M.~Gerhard, R.~Grieser, G.~Huber,
  A.~Karafillidis, M.~Krieg, N.~Schmidt, T.~K\"uhl, R.~Neumann, V.~Balykin,
  M.~Grieser, D.~Habs, E.~Jaeschke, D.~Kr\"amer, M.~Kristensen, M.~Music,
  W.~Petrich, D.~Schwalm, P.~Sigray, M.~Steck, B.~Wanner, and A.~Wolf, ``First
  laser cooling of relativistic ions in a storage ring,'' {\em Phys. Rev.
  Lett.}, vol.~64, pp.~2901--2904, Jun 1990.

\bibitem{Lauer1998}
I.~Lauer, U.~Eisenbarth, M.~Grieser, R.~Grimm, P.~Lenisa, V.~Luger,
  T.~Sch\"atz, U.~Schramm, D.~Schwalm, and M.~Weidem\"uller, ``Transverse laser
  cooling of a fast stored ion beam through dispersive coupling,'' {\em Phys.
  Rev. Lett.}, vol.~81, pp.~2052--2055, Sep 1998.

\bibitem{Hangst1991}
J.~S. Hangst, M.~Kristensen, J.~S. Nielsen, O.~Poulsen, J.~P. Schiffer, and
  P.~Shi, ``Laser cooling of a stored ion beam to 1 mk,'' {\em Phys. Rev.
  Lett.}, vol.~67, pp.~1238--1241, Sep 1991.

\bibitem{Schramm2005}
U.~Schramm, M.~Bussmann, D.~Habs, T.~Kuhl, K.~Beckert, P.~Beller, B.~Franzke,
  F.~Nolden, M.~Steck, G.~Saathoff, S.~Reinhardt, and S.~Karpuk, ``Laser
  cooling of relativistic heavy ion beams,'' in {\em Proceedings of the 2005
  Particle Accelerator Conference}, pp.~401--403, May 2005.

\bibitem{Hangst1995}
J.~S. Hangst, J.~S. Nielsen, O.~Poulsen, P.~Shi, and J.~P. Schiffer, ``Laser
  cooling of a bunched beam in a synchrotron storage ring,'' {\em Phys. Rev.
  Lett.}, vol.~74, pp.~4432--4435, May 1995.

\bibitem{Winters2015}
D.~Winters, T.~Beck, G.~Birkl, C.~Dimopoulou, V.~Hannen, T.~K{\"u}hl,
  M.~Lochmann, M.~Loeser, X.~Ma, F.~Nolden, W.~N{\"o}rtersh{\"a}user, B.~Rein,
  R.~Sánchez, U.~Schramm, M.~Siebold, P.~Spiller, M.~Steck, T.~St{\"o}hlker,
  J.~Ullmann, T.~Walther, W.~Wen, J.~Yang, D.~Zhang, and M.~Bussmann, ``Laser
  cooling of relativistic heavy-ion beams for fair,'' {\em Physica Scripta},
  no.~T166, p.~014048, 2015.

\bibitem{poth1990}
H.~Poth, ``Electron cooling: Theory, experiment, application,'' {\em Physics
  Reports}, vol.~196, no.~3, pp.~135 -- 297, 1990.

\bibitem{schramm2004}
U.~Schramm and D.~Habs, ``Crystalline ion beams,'' {\em Progress in particle
  and Nuclear Physics}, vol.~53, no.~2, pp.~583--677, 2004.

\bibitem{noda2014}
A.~Noda and et~al, ``Ultralow emittance beam production based on doppler laser
  cooling and coupling resonance,'' in {\em Proc. of International Particle
  Accelerator Conference (IPAC'14)}, International Particle Accelerator
  Conference, 2014.

\bibitem{johnson1996}
W.~Johnson, Z.~Liu, and J.~Sapirstein, ``Transition rates for lithium-like
  ions, sodium-like ions, and neutral alkali-metal atoms,'' {\em Atomic Data
  and Nuclear Data Tables}, vol.~64, no.~2, pp.~279--300, 1996.

\bibitem{metcalf1999}
H.~Metcalf and P.~Van~der Straten, {\em Laser Cooling and Trapping}.
\newblock Graduate texts in contemporary physics, Springer, 1999.

\bibitem{Schramm2004-2}
U.~Schramm, M.~Bussmann, and D.~Habs, ``From laser cooling of non-relativistic
  to relativistic ion beams,'' {\em Nuclear Instruments and Methods in Physics
  Research Section A: Accelerators, Spectrometers, Detectors and Associated
  Equipment}, vol.~532, no.~1–2, pp.~348 -- 356, 2004.
\newblock International Workshop on Beam Cooling and Related Topics.

\bibitem{Winters2013}
W.~Danyal and et~al, ``Laser cooling of relativistic $\mbox{C}^{3+}$ ion beams
  with a large initial momentum spread,'' in {\em Proceeding of Cool13},
  Cool13, 2013.

\bibitem{Beck2016}
T.~Beck, B.~Rein, F.~S\"{o}rensen, and T.~Walther, ``Solid-state-based laser
  system as a replacement for $\mbox{Ar}^+$ lasers,'' {\em Opt. Lett.},
  vol.~41, pp.~4186--4189, Sep 2016.

\bibitem{bussmann2014}
M.~Bussmann, ``Laser cooling of ion beams at relativistic energies,'' {\em ICFA
  Beam Dynamics Newsletter No. 65}, 2014.

\bibitem{Siebold2016}
M.~Siebold, M.~Loeser, F.~R{\"o}ser, D.~Albach, M.~Bussmann, S.~Eckhardt, A.~F.
  Lasagni, R.~Sauerbrey, and U.~Schramm, ``High energy yb:yag active mirror
  laser system for transform limited pulses bridging the picosecond gap,'' {\em
  Laser \& Photonics Reviews}, vol.~10, no.~4, pp.~673--680, 2016.

\bibitem{Boine2006}
O.~Boine-Frankenheim, R.~Hasse, F.~Hinterberger, A.~Lehrach, and P.~Zenkevich,
  ``Cooling equilibrium and beam loss with internal targets in high energy
  storage rings,'' {\em Nuclear Instruments and Methods in Physics Research
  Section A: Accelerators, Spectrometers, Detectors and Associated Equipment},
  vol.~560, no.~2, pp.~245 -- 255, 2006.

\bibitem{khateeb2001}
A.~M. Al-khateeb, O.~Boine-Frankenheim, I.~Hofmann, and G.~Rumolo, ``Analytical
  calculation of the longitudinal space charge and resistive wall impedances in
  a smooth cylindrical pipe,'' {\em Phys. Rev. E}, vol.~63, p.~026503, Jan
  2001.

\bibitem{Hannes1984}
H.~Risken, {\em The Fokker-Planck Equation}.
\newblock Springer, 1984.

\bibitem{Bjorken1982}
J.~D. Bjorken and S.~K. Mtingwa, ``{Intrabeam Scattering},'' {\em Part.
  Accel.}, vol.~13, pp.~115--143, 1983.

\bibitem{Lee2012}
S.~Y. Lee, {\em Accelerator Physics (Third Edition)}.
\newblock World Scientific Publishing Company, 2012.

\bibitem{bussmann2007}
M.~Bussmann, U.~Schramm, D.~Habs, M.~Steck, T.~K{\"u}hl, K.~Beckert, P.~Beller,
  B.~Franzke, W.~N{\"o}rtersh{\"a}user, C.~Geppert, {\em et~al.}, ``The
  dynamics of bunched laser-cooled ion beams at relativistic energies,'' in
  {\em Journal of Physics: Conference Series}, vol.~88, p.~012043, IOP
  Publishing, 2007.

\bibitem{betacool2008}
I.~Meshkov and et~al., {\em BETACOOL Physics Guide}, 2008.

\bibitem{BoineFrankenheim2005}
O.~Boine-Frankenheim and T.~Shukla, ``Space charge effects in bunches for
  different rf wave forms,'' {\em Physical Review Special Topics - Accelerators
  and Beams}, vol.~8, mar 2005.

\bibitem{molmer1993}
K.~M{\o}lmer, Y.~Castin, and J.~Dalibard, ``Monte carlo wave-function method in
  quantum optics,'' {\em JOSA B}, vol.~10, no.~3, pp.~524--538, 1993.

\end{thebibliography}







\end{document}